\documentclass[aps,prl,twocolumn, longbibliography,footinbib,superscriptaddress, preprintnumbers]{revtex4-1}
\usepackage[normalem]{ulem}

\usepackage{hyperref}
\usepackage{amsmath}
\usepackage{amssymb}
\usepackage{amsthm}
\usepackage{graphicx}
\usepackage{color}
\renewcommand{\thanks}[1]{\footnote{#1}}

\makeatletter
\def\@hangfrom@section#1#2#3{\@hangfrom{#1#2}#3}%\MakeTextUppercase{#3}}%
\def\@hangfroms@section#1#2{#1#2}%\MakeTextUppercase{#2}}%
\makeatother

\newcommand{\bea}{\begin{eqnarray}}
\newcommand{\eea}{\end{eqnarray}}
\newcommand{\be}{\begin{eqnarray}}
\newcommand{\ee}{\end{eqnarray}}

\def\ie{\begin{equation}\begin{aligned}}
\def\fe{\end{aligned}\end{equation}}
\def\half{{\scriptstyle \frac 12}}

\def\ie{\begin{equation}\begin{aligned}}
\def\fe{\end{aligned}\end{equation}}

\def\cD{{\cal D}}

\def\cN{{\cal N}}
\def\cO{{\cal O}}

\def\Re{{\rm Re \,}}

\def\tr{{\rm tr}}

\def\det{{\rm det \,}}

\begin{document}

\preprint{
QMUL-PH-26-11
}

\title{Giant graviton integrated correlators at finite coupling and all orders in $1/N$}
\author{Augustus Brown}
\affiliation{Centre for Theoretical Physics, Department of Physics and Astronomy, Queen Mary University of London, London, E1 4NS, UK}
\author{Daniele Dorigoni}
\affiliation{ Centre for Particle Theory \& Department of Mathematical Sciences,  Durham University, Durham DH1 3LE, UK}
  \author{Congkao Wen}
\affiliation{Centre for Theoretical Physics, Department of Physics and Astronomy, Queen Mary University of London, London, E1 4NS, UK}
\begin{abstract}

We study the giant graviton integrated correlator in SU$(N)$ $\mathcal{N}=4$ super Yang–Mills at finite complexified coupling $\tau$. Despite the formidable complexity arising from the heavy nature of the operators considered, the large-$N$ expansion simplifies dramatically and exhibits manifest modular invariance. At each order in $1/N$, the expansion coefficients are linear combinations of non-holomorphic Eisenstein series thus capturing the full spectrum of perturbative and non-perturbative effects in the Yang-Mills coupling. Furthermore, we find additional contributions, which are modular functions exponentially suppressed in $N$.
In the ’t Hooft limit, this yields an all-orders result in the $1/N$ expansion at arbitrary coupling $\lambda$, extending beyond prior results of leading orders. For the U$(N)$ theory, we obtain a closed-form expression valid for all $N$ and $\tau$, and show that the coupling-dependent sector of the large-$N$ expansion is universal between SU$(N)$ and U$(N)$ to all orders. Crucially, we exploit the integrated correlator constraints and determine the giant graviton correlator itself to two-loop order at finite $N$, previously only accessible in the planar limit.

\end{abstract}

\maketitle

There has been great interest in the study of correlation functions in $\cN=4$ supersymmetric Yang–Mills (SYM) theory. Via the AdS/CFT correspondence, these correlators are holographically dual to superstring scattering amplitudes in AdS and have provided deep insights into superstring theory in various regimes. Traditionally, such quantities have been studied perturbatively in the large ’t Hooft coupling expansion~\cite{Freedman:1998tz, Rastelli:2016nze, Aprile:2017bgs, Caron-Huot:2018kta,  Alday:2019nin, Alday:2023mvu, Aprile:2024lwy, Wang:2025pjo}, see also~\cite{Heslop:2022xgp, Bissi:2022mrs}. The introduction of integrated correlators~\cite{Binder:2019jwn, Chester:2020dja} (see \cite{Dorigoni:2022iem} for a review), combined with powerful arguments coming from S-duality~\cite{Chester:2019jas, Chester:2020vyz, Dorigoni:2021guq, Dorigoni:2021bvj, Alday:2023pet} and the mathematics of non-holomorphic modular functions, has since made it possible to access exact, non-perturbative properties of these physical observables~\cite{Chester:2019jas, Chester:2020vyz, Chester:2021aun, Caron-Huot:2024tzr, Dempsey:2025yiv}.  

In this work, we study correlators in $\cN=4$ SYM with gauge group SU$(N)$ and U$(N)$ involving two light superconformal primary operators in the stress-tensor multiplet, and two heavy operators corresponding to {sphere} 
giant gravitons, whose conformal dimensions scale as $N$~\cite{McGreevy:2000cw,Hashimoto:2000zp, Balasubramanian:2001nh, Corley:2001zk}. We refer to these quantities as giant graviton heavy–heavy–light–light (HHLL) correlators. The giant gravitons here considered are holographically dual to D3-branes wrapping part of the underlying $S^5$ geometry. One can also consider D3-branes wrapping part of the AdS$_5$ geometry as discussed in \cite{Brown:2025huy}.  These HHLL correlator can then be interpreted as scattering amplitudes of two gravitons off D3-branes, thus making these correlators of crucial importance for probing D-brane dynamics in AdS.  
{
From a quantum gravity point of view, giant gravitons probe intermediate  scales sitting between perturbative (super-)graviton states and black hole states. This fact has already been appreciated in the study of the superconformal index of $\mathcal{N}=4$ SYM which admits an expansion as a sum over sectors with different numbers of giant gravitons \cite{Imamura:2021ytr,Gaiotto:2021xce}. The fluctuations in each sector have a semi-classical interpretation on the dual gravity side, see e.g. \cite{Eleftheriou:2023jxr}.}
Furthermore, since giant gravitons are constructed as determinant operators they can be thought as being part of the baryonic sector of $\mathcal{N}=4$ SYM, closely related to baryons in large-$N$ QCD \cite{Witten:1979kh} and their geometric AdS/CFT realisation \cite{Witten:1998xy}.

 Compared to correlators of operators with fixed  dimensions, much less is known about giant gravitons. Even in free theory, the calculation of giant graviton correlators already exhibits considerable combinatorial complexity due to the large dimension of such operators (see e.g. \cite{Corley:2001zk, deMelloKoch:2004crq,Kimura:2007wy,Bissi:2011dc,Jiang:2019xdz,Jiang:2019zig, Holguin:2022zii}). Quantum corrections to giant graviton HHLL correlators have so far only been obtained in the leading large-$N$ limit, where results have been derived at weak coupling up to two loops~\cite{Jiang:2019xdz, Wu:2025ott, Jiang:2023uut}, while at strong-coupling only the leading order is known~\cite{Chen:2025yxg, Chen:2026ium}.

Thanks to a clever use of supersymmetric localization and matrix model techniques, it was shown in \cite{Brown:2024tru} that   one can go much further and obtain exact results for  integrated giant graviton correlators in the ’t Hooft large-$N$ expansion. However, unlike integrated correlators of light operators which have been solved exactly at finite complexified coupling $\tau$ and arbitrary $N$ \cite{Dorigoni:2021guq, Dorigoni:2021bvj, Dorigoni:2022cua}, the giant graviton integrated correlator has proven to be much harder, and all existing results for this quantity have so far been restricted to the ’t Hooft large-$N$ limit~\cite{Brown:2024tru, Brown:2025huy}. 

In this paper, we finally obtain exact solutions for giant graviton integrated correlators valid at generic Yang-Mills coupling $\tau$, including all perturbative and non-perturbative effects and manifesting  modular properties. Our results hold to all orders in the large-$N$ expansion for the SU$(N)$ theory, while in the U$(N)$ case we provide a {particularly} simple expression valid for generic $N$. 

What makes the final result more striking is that  instanton effects in the presence of higher-dimensional operators are not known in general, and the dimension-$N$ nature of the giant graviton means the localization  computation grows rapidly in complexity with $N$. Since our results are exact in $\tau$, they fix the ’t Hooft large-$N$ expansion to arbitrary order, extending the results of~\cite{Brown:2024tru}. Our analysis provides exact constraints for the `un-integrated' giant graviton correlator. Utilizing  these constraints, we determine the correlator up to two loops for general $N$, a quantity so far only known at large-$N$.

\vspace{-0.56cm}
\subsection{Giant graviton integrated correlators} 
\vspace{-0.2cm}

We will study the HHLL correlator given by
\begin{align} \label{eq:corrdef}
\! T(x_i,\! Y_i)\,{=}\frac{  \langle  \mathcal{D}(x_1, \!Y_1) \mathcal{D}(x_2, \!Y_2) \mathcal{O}_2(x_3,\!Y_3) \mathcal{O}_2(x_4, \!Y_4) \rangle }{ \langle  \mathcal{D}(x_1, \!Y_1) \mathcal{D}(x_2, \!Y_2) \rangle} , 
\end{align}
where $x_i$ are spacetime coordinates and $Y_i$ are SO$(6)_R$ null vectors. The giant-graviton determinant operator $\mathcal{D}$ and single-trace operators $\mathcal{O}_p$ are defined as
\begin{align} \label{eq:operators}
    \mathcal{D}(x, Y) &= \det  (\Phi_I(x) Y^I) \, ,  \\
    \mathcal{O}_p(x, Y) &=\tr (\Phi_{I_1}(x) \cdots \Phi_{I_p}(x) ) Y^{I_1} \cdots Y^{I_p} \, .
\end{align} 
The operator $\mathcal{O}_p$ has scaling dimension $\Delta_p=p$, while the operator $\mathcal{D}$ has $\Delta_{\mathcal{D}} = N$ thus becoming heavy at large-$N$. In this limit, $\mathcal{D}$ can be interpreted holographically as being dual to a D3-brane~\cite{McGreevy:2000cw,Hashimoto:2000zp, Balasubramanian:2001nh, Corley:2001zk}. Superconformal invariance~\cite{Eden:2000bk, Nirschl:2004pa} constraints \eqref{eq:corrdef} to
\begin{align}\label{eq:redCorr}
    T(x_i, Y_i) = {\rm free\,\, part} + \mathcal{I}_4(x_i, Y_i)\, \mathcal{T}_N(U, V; \tau) \, ,
\end{align}
where the prefactor $\mathcal{I}_4(x_i, Y_i)$ is fixed by kinematics. The reduced correlator $\mathcal{T}_N(U, V; \tau)$ encodes all the dynamics, depending on the rank of the gauge group $N$, the cross ratios, and  most importantly the Yang-Mills coupling
\begin{align}
    U &\label{eq:Cross}= \frac{x_{12}^2 x_{34}^2}{x_{13}^2 x_{24}^2} \, , \quad   V= \frac{x_{14}^2 x_{23}^2}{x_{13}^2 x_{24}^2} \,, \\
     \tau &= \tau_1 + {\rm i} \tau_2  = \frac{\theta}{2\pi} + {\rm i} \frac{4\pi}{g^2_{_{\rm YM}}}\, . 
\end{align}
The giant graviton \textit{integrated} correlator, $\mathcal{C}_{\mathcal{D}} (\tau; N)$, is defined by integrating out the coordinate dependence of the reduced correlator against a supersymmetry-preserving measure whose explicit form is found in~\cite{Binder:2019jwn, Chester:2020dja}. 
Importantly, this  integrated correlator can also be computed from the $S^4$ partition function of $\mathcal{N}=2^*$ SYM via 
\begin{align} \label{eq:CD-or}
\mathcal{C}_{\mathcal{D}} (\tau; N) = \frac{\partial_{\mathcal{D} } \partial_{\bar{ \mathcal{D} } }  \partial_m^2 \log Z  (\tau, \tau', m) } {\partial_{\mathcal{D} } \partial_{\bar {\mathcal{D} } }  \log Z  (\tau, \tau', m)  } \Big \vert_{\tau'=0, m=0}\, .   
\end{align}
Here $Z (\tau, \tau', m)$ denotes the $\mathcal{N}=2^*$ sphere partition function deformed by higher-dimension operators $\mathcal{O}_p$ with $p\neq2$ defined in \eqref{eq:operators} with couplings $\tau'_p$ , which we write concisely as $\tau'$ in $Z  (\tau, \tau', m)$. From supersymmetric localization~\cite{Pestun:2007rz}, this  partition function reduces to the matrix model integral,
\begin{align} \label{eq:partition}
Z  (\tau, \tau', m)  &=\! \int \! d \nu(a_i)   \, |Z_{\rm classic}(a_i; \tau, \tau')|^2 \\
& \times\,  Z_{\rm 1-loop}(a_i; m)| Z_{\rm inst}(a_i; m, \tau, \tau') |^2 \, .  \nonumber
\end{align}
The integration variables $a_i\in \mathbb{R}$ with $i=1,...,N$ are unconstrained for the U$(N)$ gauge theory, while they satisfy $ \sum_{i=1}^N a_i=0$ for SU$(N)$.
In both cases the integration measure is  $d \nu(a_i) = d^{N} \! a   \prod_{i<j} a_{ij}^2$.

To compute the giant graviton integrated correlator using localization matrix model as in \eqref{eq:CD-or} we need to insert the determinant operators via the action of $\partial_{\mathcal{D} }$ (and its conjugate $ \partial_{\bar {\mathcal{D} } } $). In the matrix model, this is realised by expressing the determinant operator in terms of trace operators of dimension $N$, i.e. ($\tr\, a^N, \tr\, a^{N-1} \tr\, a$, ...), as well as lower-dimensional operators which must be included via a Gram-Schmidt process~\cite{Gerchkovitz:2016gxx}, due to the fact that on $S^4$ operators with different dimensions mix.  Trace operators can then be expressed in terms of derivatives with respect to $\tau'$ and $\tau$. For example, for small values of $N$ in SU$(N)$ we have
\begin{align} \label{eq:tau-prime}
N&=2: \,   \partial_{\mathcal{D} } = -{1\over 2} \frac{\partial_{\tau}}{{\rm i} \, \pi } - {3 \over 4} \, ;  \quad N=3: \,   \partial_{\mathcal{D} } = {1\over 3} \frac{ \partial_{\tau'_3} }{{\rm i} \, \pi^{3\over 2}} \, ;  \cr 
N&=4: \,   \partial_{\mathcal{D}} =-\frac{1}{4} \frac{ \partial_{\tau'_4}  }{{\rm i} \, \pi^{2}}   +   \frac{1}{8}  {\partial^2_{\tau} \over ({\rm i} \, \pi)^2}- \frac{5}{16} {\partial_{\tau} \over {\rm i} \, \pi }  + \frac{75}{64} \, .  
\end{align}
Similar expressions can be obtained for U$(N)$ although we stress that in the U$(N)$ theory one must also include the operator $\mathcal{O}_1$ via $\partial_{\tau'_1}$, vanishing otherwise in SU$(N)$.
The complexity of $ \partial_{\mathcal{D}}$  grows extremely rapidly with $N$, hence imposing further obstacle to obtaining exact results in the large-$N$ expansion.

We now comment on the individual contributions appearing in \eqref{eq:partition}. The one-loop determinant $Z_{\rm 1\mbox{-}loop}$ is not affected by the insertions of higher-dimensional operators and therefore independent of  $\tau'$; the classical action $Z_{\rm classical}$ does depend on the higher-dimensional coupling $\tau'$, but in a trivial manner. Finally, the instanton contribution $Z_{\rm inst}$ \cite{Nekrasov:2002qd} may be extracted from~\cite{Gottsche:2006tn, Bonelli:2014iza}, and it depends on $\tau'$ with increasing complexity as the instanton number grows larger. Therefore,  the action of the $\tau'_p$-derivatives in $\partial_{\mathcal{D}}$, as in \eqref{eq:tau-prime},  on  $Z_{\rm classical}$ simply induces the insertion of $\tr\, a^p$; however, its action on $Z_{\rm inst}$ is highly non-trivial, becoming unyielding at large $N$. In Appendix \ref{eq:part-fun} the reader can find  expressions for $Z_{\rm 1\mbox{-}loop}$, $Z_{\rm classical}$, and the one-instanton contribution to  $Z_{\rm inst}$.

Despite these difficulties, by exploiting S-duality of $\mathcal{N}=4$ SYM~\cite{Montonen:1977sn}, we will show that not only can we obtain $\mathcal{C}_{\mathcal{D}}(\tau; N)$, but that the final result takes an {unexpectedly} simple form to all orders in the large-$N$ expansion at generic $\tau$, containing all instanton contributions. 

\vspace{-0.4cm}
\subsection{ Exact solution for the SU$(N)$ giant graviton integrated correlator} 
\vspace{-0.2cm}

For generic $N$ and $\tau$, as verified in many non-trivial examples, integrated correlators of half-BPS operators can be expressed in terms of a two-dimensional lattice-sum representation \cite{Dorigoni:2021guq, Dorigoni:2021bvj}. This formulation manifests several important properties of these correlators, in particular their modular structures. A consequence of the lattice sum representation is that the SL$(2,\mathbb{Z})$ spectral decomposition of the integrated correlators is free of Maass cusp forms and therefore takes the form \cite{Collier:2022emf,Paul:2022piq}
\begin{align} \label{eq:spectral}
 \mathcal{C}_{\mathcal{D}}(\tau; N)
 = C(N)
 + \int_{\Re\, s=\frac{1}{2}} \frac{ds}{2\pi {\rm i} } \, g_N(s) (2s{-}1)^2  E^*(s;\tau)\, ,
\end{align}
 where ${C}(N) =\lim_{s \to 1} g_N(s)$ is independent of $\tau$. The entire dependence from the coupling is carried by the non-holomorphic Eisenstein series $E^*(s; \tau)=E^*(1{-}s; \tau)$,
 \begin{align}
 &  E^*(s;\tau) 
\label{Eisen_series_definitions} =\xi(2s)\tau_2^s  +  \xi(2s-1)\, \tau_2^{1-s}\\
&\notag \phantom{=}+ \sum_{   k\neq0}e^{2\pi i k\tau_1}    2\sqrt{\tau_2}\,  |k|^{s-\half}\sigma_{1-2s} (k)     \, K_{s-\half}(2\pi |k| \tau_2) \,,
\end{align}
where $\xi(s) = \pi^{-s/2}\Gamma(\frac{s}{2})\zeta(s)=\xi(1{-}s)$ is the completed zeta function, $\sigma_s(n) = \sum_{d\vert n }d^s$ is the divisor sigma function, and $K_s(z)$ denotes a modified Bessel function of the second kind.
The $k^{th}$ Fourier-mode corresponds to a $k$-instanton contribution. The expression \eqref{eq:spectral} makes manifest the SL$(2, \mathbb{Z})$-invariance and contains all the perturbative and instanton contributions. Importantly, the \textit{spectral overlap} $g_N(s)$ is reflection invariant, i.e. $g_N(s)=g_N(1{-}s)$, and can be determined by comparing \eqref{eq:spectral} with the purely perturbative expansion of the integrated correlator obtained from the supersymmetric localization matrix model, which takes the form  
        \begin{align} \label{eq:pert-2loop}
\mathcal{C}_{\mathcal{D}} (\tau; N)\vert_{\rm pert} &=     N    \frac {3 \zeta (3) \lambda} {2\pi^2}  -  \bigg[ \frac{ 3 N (N {+} 1)^2 + 2 }{N (N+1)}  \\ 
+ & \frac{(N^2{-}1) \left(N+2\right)}{N (1 {-} (-N)^{N+1})}\bigg] \frac{15 \, \zeta (5) \lambda^2}{ 32 \, \pi^4} + \ldots \, , \nonumber 
  \end{align}
where $\lambda=N g^2_{_{\rm YM}}$. 
We see that the two-loop result naturally splits into two parts, one of which is proportional to $1/(1 - (-N)^{N+1})$ and is therefore exponentially suppressed in the large-$N$ expansion. We have verified that the same structure persists at higher orders~\footnote{{It can be seen from the matrix model through the small coupling expansion of $Z_{1-\text{loop}}$ that the $O(\lambda^1)$ term contains only $\tr\, a^2$, which implies that this contribution can be obtained by $\tau_2 \partial_{\tau_2}$ acting on the two-point function of giant gravitons (then normalized by the two-point function itself), so the complicated color factors arising from giant gravitons cancel out. Therefore there is no exponentially suppressed contribution for the $\zeta(3)$ term in \eqref{eq:pert-2loop}.}}.

In view of this structure \eqref{eq:pert-2loop} (and its higher-order analogy), we express the spectral overlap as the sum of two contributions,
  \begin{align} \label{eq:gN(s)}
 g_N(s)  = g^{(1)}_N(s)  + g^{(2)}_N(s)   \, . 
  \end{align}
Using \eqref{Eisen_series_definitions}, we evaluate the perturbative expansion of the spectral integral \eqref{eq:spectral} and identify
 $g^{(1)}_N(s)$ and $g^{(2)}_N(s)$ respectively as the parts of the spectral overlap responsible for the rational contribution in $N$ and the term proportional to  $1/(1 {-} (-N)^{N+1})$ in the expansion \eqref{eq:pert-2loop}.

By exploiting the matrix model \eqref{eq:partition}, we obtain an exact result for $ g^{(1)}_N(s)=f_N(s) + h_N(s)$ where 
    \begin{align} \label{eq:fN}
 f_N(s) &={N\, \pi \over \sin(\pi s)}  \, _3F_2(1-N,1-s,s;2,2;1)   \, , 
\\ 
h_N(s) &\notag =  \frac{  \pi\, _2F_1\left(-N{-}1,-s;-N{-}s;-\frac{1}{N}\right) { N+s \choose s}}{\sin(\pi s) \, (1-2s)  \left(1+ {1\over N} \right)^s} \\
&\phantom{=}+(s\leftrightarrow 1-s)\, . \label{eq:hN}
  \end{align}
Note that $f_N(s) = f_N(1-s)$ so that the partial spectral overlap is reflection invariant $g^{(1)}_N(s) = g^{(1)}_N(1-s)$.
For $s\in \mathbb{N}$, we find that $g^{(2)}_N(s)$  has the following structure, 
      \begin{align}
  g^{(2)}_N(s) =  \frac{(N{-}1)  (N{+}2) (N{+}1)^{4-s}}{1 {-} (-N)^{N+1}} \sum_{i=0}^{3s-9} c_i (s) \, N^i \, ,
  \end{align}
where $g^{(2)}_N(2)=0$. While it is straightforward to compute $g^{(2)}_N(s)$ for a given $s\in \mathbb{N}$, a general closed-form expression remains out of reach. Nevertheless, the key observation is that $g^{(2)}_N(s)$ is responsible for producing the factor $1/(1 {-} (-N)^{N+1})$ as in \eqref{eq:pert-2loop}, rendering it exponentially suppressed at large $N$ -- precisely the regime relevant for giant gravitons analysed in the next section.

 We have verified \eqref{eq:spectral} through 12 loops in perturbation theory  using $Z_{\rm 1-loop}$ via the `full algebra method' \cite{Billo:2017glv}, and at the one-instanton level for all $N \leq 10$ using the one-instanton partition function which can be found in Appendix \ref{eq:part-fun}. 

As emphasized, the result imposes exact integral constraints on the giant graviton correlator. Applying these using the perturbative expansion \eqref{eq:pert-2loop}, together with additional OPE constraints {coming from the LL and HH channels}, we determine the reduced correlator to two-loop order for general $N$~\footnote{See \cite{Caron-Huot:2023wdh} for the study of determinant operators related to the one here considered at finite $N$ from a twistor space approach.}
\begin{align}
&\label{eq:2loop} \mathcal{T}_N(U, V; \tau) |_{\rm pert}=N x_{13}^4 x_{24}^4 \, \bigg[\frac{\lambda}{4\pi^2} f^{(1)}(x_i)  \\ 
 - &  \frac{1}{8}\! \left( {\lambda \over 4\pi^2} \right)^2 \!\! \left(  f^{(2)}_{2}(x_i) {-} f^{(2)}_{1}(x_i)  {+}   c(N) f^{(2)}_{1}(x_i)\right) \bigg]{+}\, O(\lambda^3)\,, \nonumber
\end{align}
where the color factor $c(N)$ contains the non-planar and exponentially suppressed contributions
\begin{align} \label{eq:two-loops-corr2}
    c(N) = \frac{3 N (N+1)+2}{ N^2 (N+1)}+  \frac{ (N^2-1) (N+2)}{N^2(1- (-N)^{N+1})} \, .
\end{align}
The  $f^{(\ell)}_{\alpha}(x_i)$ are $f$-graph functions introduced in~\cite{Eden:2011we, Eden:2012tu}. In Appendix \ref{app:twoloops}, the reader can find explicit expressions for the one-loop box integral $f^{(1)}(x_i)$, and the two-loop double-box and squared one-loop box integrals contained in  $f^{(2)}_{1}(x_i)$ and $f^{(2)}_{2}(x_i)$. 
 At large-$N$, $c(N)$ is suppressed and \eqref{eq:2loop} agrees with  \cite{Jiang:2019xdz, Jiang:2023uut}. Furthermore, for $N=2$ and $N= 3$ the result reproduces the known correlators $  \langle \mathcal{O}_2 
\mathcal{O}_2  \mathcal{O}_2   \mathcal{O}_2  \rangle$ for SU$(2)$ and $    \langle \mathcal{O}_3 
\mathcal{O}_3  \mathcal{O}_2  \mathcal{O}_2   \rangle$ for SU$(3)$ \cite{Eden:2012tu, Chicherin:2015edu}. 
We refer to Appendix \ref{app:twoloops} for more details.

\vspace{-0.4cm}
\subsection{Giant graviton integrated correlator at large-$N$}
\vspace{-0.2cm}

We now derive the large-$N$ expansion of the integrated correlator at fixed coupling $\tau$. As discussed, $g^{(2)}_N(s)$ in \eqref{eq:gN(s)} is exponentially suppressed and we will omit its contribution in this limit.  
Hence we shall focus our attention solely towards $g^{(1)}_N(s)$. Using \eqref{eq:fN}-\eqref{eq:hN} we compute the large-$N$ expansions for the building blocks \allowdisplaybreaks{
\begin{align}
f_N(s) &\label{eq:largeNspec1}= \sum^{\infty}_{\ell=0} N^{1-s-\ell} f^{(\ell)}(s) +(s\to1-s)\,,\\
h_N(s) &\label{eq:largeNspec2}= \sum^{\infty}_{\ell=0} N^{1-s-\ell} h^{(\ell)}(s) +(s\to1-s)\, ,
\end{align}
where we find
\begin{align}
   f^{(\ell)}(s) &\label{eq:fell}= 
  \frac{ \tan (\pi  s) \, \Gamma (s-1)^2 }{2^{2s+\left\lfloor \frac{\ell}{2}\right\rfloor }\sqrt{\pi }\, \Gamma \left(s {+} \left\lfloor \frac{\ell}{2}\right\rfloor {+} \frac{1}{2}\right)} P^{(\ell)}_1(s)\,,\\
    h^{(\ell)}(s)
    &\label{eq:hell} =
   \frac{ \Gamma(s-1)}{(1-2s)} P^{(\ell)}_2(s) \, , 
\end{align} }
with $P_1^{(\ell)}(s),P_2^{(\ell)}(s)$ simple polynomials in $s$, e.g. 
\begin{align}
P_1^{(0)}(s) &\label{eq:P1}= 1\,,\qquad P_1^{(1)}(s) = \frac{1-s}{2}\,,\\
P_2^{(0)}(s) &\label{eq:P2}=1\,,\qquad P_2^{(1)}(s) = \frac{ (s-1) (s-2)}{2}\,.
\end{align}
Note that since the spectral integral \eqref{eq:spectral} and the Eisenstein series \eqref{Eisen_series_definitions} are invariant under $s\leftrightarrow 1-s$, we will omit the reflected part $(s\to 1-s)$ in \eqref{eq:largeNspec1}-\eqref{eq:largeNspec2} and multiply the terms presented by a factor of $2$.

Substituting the large-$N$ expansions \eqref{eq:largeNspec1}-\eqref{eq:largeNspec2} in \eqref{eq:spectral}, we evaluate the spectral integral by closing the contour of integration towards $\Re s\to +\infty$ while collecting residues using the expressions \eqref{eq:fell}-\eqref{eq:hell}.
We obtain in this way the modular invariant large-$N$ expansion for the giant graviton integrated correlator,
\begin{equation}
\mathcal{C}_\mathcal{D}(\tau;N) \sim \widetilde{{C}}(N) -\widehat{E}(1; \tau) - \sum_{\ell=1}^{\infty} N^{{1 \over 2}-\ell} \, \mathcal{C}_{\cD}^{(\ell)}(\tau)\, . \label{eq:genusMod}
\end{equation}
Some further comments on the derivation of the above result are in order.
Firstly, at any given order in large-$N$ we find that the spectral overlap has finitely many poles located at half-integer values of $s$.
This leads to expansion coefficients which are linear combinations of Eisenstein series with half-integer indices,
\begin{align} \label{eq:lDD}
\mathcal{C}_{\cD}^{(\ell)}(\tau) = \sum_{k=1}^{\ell} c^{(\ell)}_{k}\, E\big(k+ \tfrac{1}{2}; \tau \big) \, , 
\end{align}
where  
$ E(s;\tau) = 2E^*(s;\tau)/\Gamma(s) $, 
with $E^*(s;\tau)$ given in \eqref{Eisen_series_definitions}. 
The coefficients $c^{(\ell)}_{k}$ are rational numbers, with examples given in Appendix \ref{sec:Hooft}. 

The residue at $s=1$ has to be treated carefully. For $\ell=0$, we find a double pole at $s=1$ which leads to the term $\hat{E}(1;\tau)$ in \eqref{eq:genusMod}, which is a regularised version of the singular Eisenstein series $E(1;\tau)$, i.e.
\begin{align} \label{eq:E1}
&\notag \hat{E}(1;\tau) = \lim_{s\to 1} \left[E(s;\tau) - \frac{1}{s-1}\right] \\
&= \frac{\pi \tau_2}{3} +2\gamma -\log(4\pi \tau_2)
 + 2\sum_{k\neq 0}  e^{2\pi i k\tau_1}\sum_{m|k} {1 \over m}\,,
\end{align}
where $\gamma$ is Euler-Mascheroni constant. For $\ell\neq 0$, the residue at $s=1$ yields a coupling-independent constant
which, when combined with $C(N)$ in \eqref{eq:spectral},
leads to the large-$N$ expansion of the coupling-independent part,
\begin{align}
\!\! \widetilde{{C}}(N) =  2N {-} \log(N) - 2 +\! \sum_{k=1}^{\infty}  {2 \over N^k } \!  \left[\frac{B_k}{2k}- (-1)^k\right] \, , 
\end{align}
 where $B_k$ denotes the $k$-th Bernoulli number. 
 
{These results have important holographic implications.} At large-$N$ and fixed $\tau$, since $1/\sqrt{N} \sim \alpha'$, each term appearing in \eqref{eq:genusMod} corresponds to stringy corrections to two-graviton scattering in the presence of D3-branes in AdS$_5 \times S^5$, {and provides constraints on the amplitudes.} In particular, the functions $\hat{E}(1;\tau)$ and ${E}(\tfrac{3}{2};\tau)$ are associated with the BPS higher-derivative terms $R^2, D^2R^2$, and are in agreement with {EFT} expectations {due to supersymmetry}~\cite{Bachas:1999um, Basu:2008gt, Lin:2015ixa}.  The leading term $2N$ of the coupling independent part $\widetilde{{C}}(N)$  reproduces the tree-level supergravity contribution~\cite{Chen:2025yxg}, while the subleading terms warrant further holographic study.  

We stress that the spectral integral \eqref{eq:spectral} receives additional modular invariant contributions from the large-$s$ region, which are however exponentially suppressed in $N$~\cite{Dorigoni:2024dhy}. We find two different types of such non-perturbative corrections, which at leading order are 
\begin{equation}\label{eq:NP}
\mathcal{C}_\mathcal{D}(\tau;N)\vert_{\rm NP} {=}\! -2 N^{\tfrac{3}{4}} \pi^{\frac{1}{2}} D_{\frac{N}{4}}(\tfrac{1}{4};\tau) {\pm }i 4 N^{\frac{1}{2}}  D_N(\tfrac{1}{2};\tau)+...\,.
\end{equation}
The real-analytic modular function $D_N(s;\tau)$ first introduced in \cite{Dorigoni:2022cua} (see also \cite{Luo:2022tqy}) is defined as
\begin{equation}\label{eq:DN}
D_N(s;\tau) = \!\! \! \sum_{(n,m)\neq (0,0)} \!\!\!  e^{-4\sqrt{N \pi \frac{|n\tau+m|^2}{\tau_2}}} \frac{\tau_2^s/(16\pi)^s}{ |n\tau+m|^{2s}}\,.
\end{equation}
The non-perturbative terms in \eqref{eq:NP} proportional to $D_{N/4}(s;\tau)$ give rise to the leading corrections, and they originate from the large-$s$ asymptotic of the spectral integral for $h_N(s)$ as in \eqref{eq:hell}. The sub-leading corrections in \eqref{eq:NP} proportional to $D_{N}(s;\tau)$ instead come from $f_N(s)$ as in \eqref{eq:fell}. 
In Appendix \ref{sec:Hooft} we give higher-order corrections to \eqref{eq:NP}. The seeming $\pm i$ ambiguity in \eqref{eq:NP} reflects the non-Borel summability of the asymptotic expansion \eqref{eq:genusMod}, and the non-perturbative corrections \eqref{eq:NP} are precisely the ambiguity-cancelling contributions required for a well-defined resurgent completion~\cite{Dorigoni:2024dhy}.

The usual ’t Hooft large-$N$ expansion at fixed $\lambda$ is readily extracted from \eqref{eq:genusMod} by using $\tau_2 = 4 \pi  N/\lambda$ and keeping only the zero mode of the non-holomorphic Eisenstein series \eqref{Eisen_series_definitions}. 
Explicitly, in the large-$N$ 't Hooft regime the giant graviton integrated correlator is written as
\begin{equation}
\mathcal{C}_{\mathcal{D}}(\tau; N) \sim\mathcal{I}(\lambda) =  \sum_{\ell=0}^\infty N^{1-\ell} \mathcal{I}^{(\ell)}(\lambda)\,,
\end{equation}
where the genus-$\ell$ contribution is given by
\begin{equation} \label{eq:Ilambda}
\! \mathcal{I}^{(\ell)}(\lambda)= \int_{\mathcal{C}}  \frac{{\rm d} s}{\pi i} \, {(2s-1)^2} \big[f^{(\ell)}(s) + h^{(\ell)}(s) \big] \, \xi(2s)\, \left( \frac{\lambda}{4\pi^2} \right)^{-s}  . 
\end{equation}
The contour of integration $\mathcal{C}$ is chosen as to separate the poles located at $-s\in \mathbb{N}^{\geq 1}$, whose contributions correspond to the weak coupling expansion, from all remaining poles, located at $s=0$ and $s = k+1/2$ with $k\in \mathbb{Z}$ and $k \geq -\lfloor \frac{\ell}{2}\rfloor$, which instead contribute to the strong coupling expansion.
The expression \eqref{eq:Ilambda} provides the 't Hooft large-$N$ expansion of the giant graviton integrated correlator to all orders. In particular, we reproduce the first two orders obtained in \cite{Brown:2024tru}, namely terms with $\ell=0, 1$. In Appendix \ref{sec:Hooft} we present several new higher-order terms. 

Importantly, from \eqref{eq:hell} we note that the integrand of \eqref{eq:Ilambda} proportional to $h^{(\ell)}(s)$ is analytic for ${\Re}(s)>2$. Hence for strong 't Hooft coupling, the corresponding contribution to the Mellin representation \eqref{eq:Ilambda} will produce only a finite number of $1/\lambda$ corrections and it will be otherwise responsible for the leading non-perturbative corrections of the form $O(e^{ - \sqrt{\lambda}})$.
These corrections have been studied at genus-zero in \cite{Brown:2025huy} for a more general class of giant graviton correlators. 

 Such non-perturbative effects arise precisely by considering the 't Hooft limit of the modular invariant functions $D_{N/4}(s;\tau)$ appearing in \eqref{eq:NP}.  The integrand of \eqref{eq:Ilambda} proportional to $f^{(\ell)}(s)$ is instead responsible for smaller non-perturbative corrections of the form $O(e^{ - 2\sqrt{\lambda}})$ which coincide with the 't Hooft limit of the modular invariant functions $D_{N}(s;\tau)$ appearing in \eqref{eq:NP}. Holographically, these contributions appear to be releated to $(p, q)$-string world-sheet instantons~\cite{Dorigoni:2022cua}. 
\vspace{-0.46cm}
\subsection{Exact solution for the U$(N)$ giant graviton integrated correlator}
\vspace{-0.2cm}

We now consider the  $\mathrm{U}(N)$ theory. Unlike what we found for the SU$(N)$ theory, the U$(N)$ giant graviton integrated correlator does not receive $O(N^{-N})$-type contributions. This absence is what allows us to obtain an exact closed-form solution valid for all $N$ and $\tau$. Following the same procedure carried out for SU$(N$),  we derive a spectral representation identical to \eqref{eq:spectral}, with spectral overlap given by 
\begin{align}\label{eq:UNg}
 \tilde{g}_N(s) =
f_N(s)
+ \frac{\pi \big[ \binom{N+1-s}{1-s} - \binom{N+s}{s} \big]}{(2s{-}1)\, \sin(\pi s)} \, , 
\end{align}
where $f_N(s)$ was given in \eqref{eq:fN} and we note that the second factor is again symmetric in $s\leftrightarrow 1-s$.
This simple expression provides the complete result for the giant graviton integrated correlator in the U$(N)$ theory, and we have explicitly verified the expression perturbatively and at the one-instanton level. 

The large-$N$ expansion for the second contribution to \eqref{eq:UNg} is easy to derive,
\begin{align}
\frac{\pi  \binom{N+1-s}{1-s}}{(2s{-}1)\, \sin(\pi s)} &\label{eq:largeNUN} = \sum^{\infty}_{\ell=0} N^{1-s-\ell} 
   \frac{ \Gamma(s-1)}{(1-2s)} \tilde{P}^{(\ell)}(s) \, , 
\end{align}
and it is precisely of the same form as that of $h_N(s)$ given in \eqref{eq:largeNspec2}-\eqref{eq:hell}.
The polynomials $\tilde{P}^{(\ell)}(s)$ with $\ell=0,1$ in fact coincide with the corresponding SU$(N)$ results \eqref{eq:P2}, and furthermore, just as for the expansion of $h_N(s)$ in SU$(N)$, we note that at large-$N$ each order of \eqref{eq:largeNUN} is analytic for ${\rm Re}(s)>1$.
As explained below \eqref{eq:genusMod}, from the large-$N$ expansion of \eqref{eq:UNg} we evaluate the analogous spectral overlap integral \eqref{eq:spectral} for the U$(N)$ giant graviton correlator by closing the $s$-contour of integration to ${\rm Re}(s)\to +\infty$.
As a consequence of these facts, we find that at large-$N$ the leading and sub-leading orders of the SU$(N)$ and U$(N)$ theories are identical, in agreement with the results of~\cite{Brown:2024tru} in the large-$N$ 't Hooft expansion.  At higher orders, the polynomials $\tilde{P}^{(\ell)}(s)$ generally differ between the two theories; however, these differences  affect only the coupling-independent constants and exponentially suppressed terms.

In terms of 't Hooft expansions, this means that at any given genus the SU$(N)$ and U$(N)$ giant graviton integrated correlators are identical to all orders in the strong-coupling expansion, up to $\lambda$-independent constants. This suggests that certain coupling-dependent dynamics of giant graviton correlators are insensitive to the difference between SU$(N)$ and U$(N)$ gauge groups, and constitutes another universal property of giant graviton integrated correlators, analogous to those uncovered in~\cite{Brown:2025huy, DeLillo:2025stg}.

\vspace{-0.6cm}
\subsection{Conclusion and discussion}
\vspace{-0.3cm}

In this work, we obtain exact solutions in the Yang-Mills coupling $\tau$ for the giant graviton integrated correlator in $\cN=4$ SYM, to all orders in the large-$N$ expansion for gauge group SU$(N)$  and for arbitrary $N$ in U$(N)$,
{via the integral-representation \eqref{eq:spectral} with spectral overlaps given in equations \eqref{eq:fN}-\eqref{eq:hN} for SU$(N)$ and in \eqref{eq:UNg} for U$(N)$}. Despite the formidable complexity of the underlying computation, the result takes a remarkably simple form, making SL$(2,\mathbb{Z})$
invariance manifest. 
Thanks to our results, we determine at weak coupling the giant graviton HHLL correlator to the two-loop order for arbitrary $N$. 
We also reveal that the coupling-dependent part of the large-$N$ strong 't Hooft coupling expansion is universal between SU$(N)$ and U$(N)$ theories to all orders. Holographically, our large-$N$ analysis at finite $\tau$ provides exact constraints on two-graviton/D3-brane scattering in AdS$_5 \times S^5$ at finite string coupling. 

Our results open up several interesting directions for future work. An immediate question is to determine the $O(N^{-N})$ contributions in the SU$(N)$ case and, perhaps more importantly, to understand their holographic interpretation. {This may be achievable by finding certain recursion relations like Laplace-difference equations of~\cite{Dorigoni:2021bvj, Dorigoni:2021guq}, which would also provide a pathway to rigorously prove the results for any number of instantons following ideas of~\cite{Dorigoni:2022cua}.} Another direction is to extend our analysis to sub-determinant operators and dual giant gravitons for which presently only the leading large-$N$ ’t Hooft expansion is known \cite{Brown:2025huy}, and the associated modular functions have yet to be understood. {These results would also allow for studying the intriguing relation between sphere giant gravitons and AdS giant gravitons via the map $N \leftrightarrow -N$, as found in~\cite{Brown:2025huy}, beyond the planar limit~\footnote{{Preliminary calculations for general sphere giant gravitons with dimension $\alpha N$ and AdS giant gravitons with dimension $\beta N$ suggest that the spectral overlap can again be separated into two parts for SU$(N)$ theory. The first part, valid at all orders in $1/N$, relates the sphere giant gravitons and AdS giant gravitons by $N\leftrightarrow -N$ at a given $s$, while the exponentially suppressed parts are not related to each other in this simple manner. For U$(N)$ theory, there are no $O(N^{-N})$ constributions, and the spectral overlap for sphere giant gravitons is fully related to the AdS giant graviton spectral overlap by $N \leftrightarrow-N$.}}.} Moreover, our all-orders results make it possible to investigate whether the universality between the giant graviton integrated correlators here discussed and those in certain $\mathcal{N}=2$ SCFTs~\cite{DeLillo:2025stg, Chester:2025ssu}, currently established at the first two orders at large $N$, extends to higher orders. Finally, it would be interesting to generalise our results to other gauge groups, where the action of S-duality on the correlators is even more nontrivial \cite{Goddard:1976qe}. In these cases, analogous lattice-sum representations have been identified only for integrated correlators of light operators \cite{Dorigoni:2022zcr, Dorigoni:2023ezg}, and it would be valuable to extend this analysis to include giant gravitons. 

\vspace{-0.58cm}
\section*{Acknowledgments}
\vspace{-0.2cm}
{\small 
We would like to thank  Mikhail Bershtein, Giulio Bonelli, Frank Coronado, Stefano Cremonesi, Francesco Galvagno, Song He,  Masazumi Honda, Sanjaye Ramgoolam, Canxin Shi, Yichao Tang, Alessandro Tanzini, and Mitchell Woolley for insightful discussions.  
   DD is supported by
the Royal Society grants ICA$\backslash$R2$\backslash$242058 and IEC$\backslash$R3$\backslash$243103. CW is supported by a Royal Society University Research Fellowship No. URF$\backslash$R$\backslash$221015 and a STFC Consolidated Grant, ST$\backslash$X00063X$\backslash$1 ``Amplitudes, strings \& duality". 

\onecolumngrid
\setcounter{secnumdepth}{2}

\appendix

\section{Sphere partition function for $\mathcal{N}=2^*$ SYM deformed by higher-dimensional operators} \label{eq:part-fun}

Here we review each contribution to the sphere partition function of $\mathcal{N}=2^*$ SYM deformed by higher-dimensional operators, given in equation (8) of the main work. For the classical part of the action, we have explicitly
\begin{align} \label{eq:Z-classical}
    Z_{\rm classic}(a_i; \tau, \tau') &=\exp \bigg[ {\rm i}\, \pi \tau \, \tr\, a^2 + \sum^\infty_{\substack{p=1 \\ p\neq 2}} {\rm i} \, \pi^{\frac{p}{2}} \tau'_p \, \tr\, a^p \bigg]\, .
    \end{align}
We see that $\tau'_1$ is only present for the U$(N)$ theory, since in SU$(N)$ we have $\tr\, a =0$ . The perturbative one-loop determinant contribution is contained in $Z_{\rm 1-loop}(a_i; m)$ and it is given by
    \begin{align} \label{eq:Z1loop}
 Z_{\rm 1-loop}(a_i; m) = \frac{1}{H^N(m)} \prod_{i<j} \frac{H^2(a_{ij})}{H(a_{ij}+m)H(a_{ij}-m)} \, ,
\end{align}
with $H(x)=e^{-(1+\gamma)x^2} G(1+{\rm i} x)\, G(1-{\rm i} x)$, where $G$ denotes the Barnes $G$-function. 
The instanton contributions, localised at the north and south poles of the 4-sphere, are affected by the higher-dimensional operators. For the one-instanton sector, it is given as
\begin{align} \label{eq:Z-oneinst}
& Z_{\rm 1-inst}(a_i; m, \tau, \tau') = -m^2 \sum_{\ell=1}^N \left( \prod_{j \neq \ell} \frac{(a_{\ell j} + {\rm i})^2 -m ^2}{(a_{\ell j} + {\rm i})^2 +1} \right)   \exp \bigg[\! -{\rm i} \sum_{p = 3}^\infty \pi^{\frac{p}{2}} \tau'_p\, (a_{\ell}^p +(a_{\ell} +2 {\rm i})^p -2(a_{\ell} + {\rm i})^p) \bigg] \, . 
\end{align}
The above expression applies to both SU$(N)$ and U$(N)$ theories, depending on whether we impose the traceless condition $\tr\, a=0$. These expressions for $ Z_{\rm 1-loop}(a_i; m) $ and $Z_{\rm 1-inst}(a_i; m, \tau, \tau')$ have been used in the main text to explicitly verify the perturbative and the one-instanton contributions of the exact expressions for the giant graviton integrated correlators. 

For generic instanton number, one may extract the instanton partition function from the results of~\cite{Gottsche:2006tn, Bonelli:2014iza}. However as argued in the main text, by exploiting modular invariance and the lattice-sum representation we have shown that the integrated correlator is fully determined by the zero-instanton purely perturbative contributions. The instanton contributions thus serve as non-trivial checks. In fact, as a consequence of modular invariance we have that since our expressions have been verified at the one-instanton (and perturbative) level, the spectral representation guarantees that our result is correct for all instanton sectors.

\section{Further results for the large-$N$ expansion} 
\label{sec:Hooft}

As showed in the main text, the integrated giant graviton correlator expanded at large-$N$ and fixed Yang-Mills coupling $\tau$ takes the following form 
\begin{align} \label{eq:applargeN1}
    \mathcal{C}_\mathcal{D}(\tau; N)& \sim \widetilde{C}(N) -\widehat{E}(1; \tau) - \sum_{\ell=1}^{\infty} N^{{1 \over 2}-\ell}  \sum_{k=1}^{\ell} c^{(\ell)}_{k}\, E\big(k+ \tfrac{1}{2}; \tau \big)\, .
    \end{align}
Here we use a slightly different normalisation for the non-holomorphic Eisenstein series where $ E(s;\tau) = 2E^*(s;\tau)/\Gamma(s)$, and $\widehat{E}(1; \tau)$ denotes the regularised ${E}(1; \tau)$, given in equation (27) of the main work. As discussed in the main text,  it is straightforward to compute the large-$N$ expansion to any orders. For example, up to $\ell=4$ we find
\begin{align} \label{eq:clk}
    \Big\{  c^{(1)}_{1} &= {1 \over 2} \Big\} \, ; \qquad   \Big\{c^{(2)}_{1} = {3 \over 2^5} \, , \quad  c^{(2)}_{2} = - {1 \over 2^3}  \Big\} ; \qquad  \Big\{ c^{(3)}_{1} = {405 \over 2^{12}} \, , \quad  c^{(3)}_{2} = - {9 \over 2^7}\, , \quad  c^{(3)}_{3} =  {25 \over 2^{10}} \Big\} ; \\ 
  \Big\{c^{(4)}_{1} &= {7875 \over 2^{15}} \, , \quad   c^{(4)}_{2} = - {2025 \over 2^{14}}\, , \quad  c^{(4)}_{3} =  {105 \over 2^{13}} \, , \quad  c^{(4)}_{4} =  {35 \over 2^{12}} \Big\}\, . \nonumber
\end{align}
It is important to note that the perturbative expansions given in \eqref{eq:applargeN1} is an asymptotic, factorially divergent power series in $1/N$ which is not Borel summable. Through resurgence analysis (see e.g. \cite{Dorigoni:2014hea} for introductory notes) this implies the presence of exponentially suppressed terms, as we will discuss below. 

On top of the perturbative expansion \eqref{eq:applargeN1}, we also find an infinite number of non-perturbative, modular invariant corrections which we denote by $ \mathcal{C}_\mathcal{D}(\tau; N)\vert_{{\rm NP},f}$ and  $ \mathcal{C}_\mathcal{D}(\tau; N)\vert_{{\rm NP},h}$ originating from the respective spectral overlaps contributions given in equations (19) and (20) of the main work.
The non-perturbative contributions produced by the large-$N$ expansion of the first spectral overlap contribution take the form 
\begin{equation}\label{eq:NPDf}
 \mathcal{C}_\mathcal{D}(\tau; N)\vert_{{\rm NP},f} = \pm i \, \sum_{\ell=0}^\infty N^{\frac{1}{2}-\ell} \sum_{k=0}^{2\ell} d_k^{(\ell)} D_N(k+\tfrac{1-3\ell}{2};\tau)\,,
\end{equation}
where the modular invariant function $D_N(s;\tau)$ has been defined in equation (30) of the main paper.
The first few coefficients are
\begin{align}
&\Big\{ d_0^{(0)} = 4 \Big\}\,,\qquad \Big\{d_0^{(1)}= -\frac{1}{96}\,,\quad d_1^{(1)} = -1\,,\quad d_2^{(1)} = 6 \Big\}\,,\\
&\notag \Big\{ d_0^{(2)} = \frac{1}{73728}\,,\quad d_1^{(2)} =\frac{1}{384} \,,\quad d_2^{(2)} =\frac{5}{64} \,,\quad d_3^{(2)} = -\frac
{1}{2}\,,\quad d_4^{(2)} =\frac{5}{2} \Big\}\,. 
\end{align}
As explained in the main text, the seeming $\pm i$ ambiguity in \eqref{eq:NPDf} is related to the non-Borel summability of \eqref{eq:applargeN1}. 

The non-perturbative contributions produced by the large-$N$ expansion of  the second spectral overlap contribution take instead the form
\begin{equation}\label{eq:NPDh}
 \mathcal{C}_\mathcal{D}(\tau; N)\vert_{{\rm NP},h} = \sum_{\ell=0}^\infty N^{\frac{3}{4}-\frac{\ell}{2}} \sum_{k=0}^\infty \tilde{d}_k^{(\ell)} D_{\frac{N}{4}}\left( \tfrac{2\ell+1}{4} {-} k ;\tau \right)\,.
\end{equation}
The coefficients $\tilde{d}^{(\ell)}_0$, $\tilde{d}^{(\ell)}_1$ for the leading and sub-leading index  $D_{N/4}(s;\tau)$ function are found to be
\begin{equation}\label{eq:dtilde}
\tilde{d}^{(\ell)}_0 = \frac{(-1)^{\ell+1} \left(4 \ell^2+3\right) \Gamma \left(\ell-\frac{3}{2}\right) \Gamma \left(\ell+\frac{1}{2}\right)}{\sqrt{\pi } \Gamma (\ell+1)}\,, \qquad \tilde{d}^{(\ell)}_1 =\frac{(-1)^{\ell+1 } \left[16 \ell^4-224 \ell^3+664 \ell^2-424 \ell+9 \right]  \Gamma \left(\ell-\frac{3}{2}\right)^2}{128 \sqrt{\pi } \Gamma (\ell+1)}\,.
\end{equation}

From the SL$(2, \mathbb{Z})$-invariant results, it is straightforward to extract the 't Hooft large-$N$ expansion. The large-$N$ expansion with fixed 't Hooft coupling $\lambda$ can be expressed as 
\begin{equation}
\mathcal{C}_{\mathcal{D}}(\tau; N) \sim\mathcal{I}(\lambda) =  \sum_{\ell=0}^\infty N^{1-\ell} \mathcal{I}^{(\ell)}(\lambda)\,,
\end{equation}
where the genus-$\ell$ contribution admits the Mellin integral representation
\begin{equation} \label{eq:IMell}
\mathcal{I}^{(\ell)}(\lambda)= \int_{\mathcal{C}}  \frac{{\rm d} s}{\pi i} \, {(2s-1)^2} \big[f^{(\ell)}(s) + h^{(\ell)}(s) \big] \, \xi(2s)\, \left( \frac{\lambda}{4\pi^2} \right)^{-s} \,,
\end{equation}
with $\xi(s) = \pi^{-s/2}\Gamma(\frac{s}{2})\zeta(s)=\xi(1-s)$. The choice of the integration contour $\mathcal{C}$ is described in details in the main work. 
Using the expressions for $f^{(\ell)}(s)$ and $h^{(\ell)}(s)$ given in equations (14) and (15) of the main article, we may derive the large-$N$ expansion with generic fixed $\lambda$ to any given orders.  For the first two orders, $O(N)$ and $O(N^0)$ we reproduce the known results in  \cite{Brown:2024tru}, obtained through a different approach. Below we present the next few orders which are novel expressions in the strong coupling expansion: 
\begin{align}
     \mathcal{I}^{(2)}  (\lambda) &\sim -\frac{\sqrt{\lambda }}{6} + \frac{3}{2} + \sum_{n=1}^{\infty} \frac{n^2 (2 n{-}7)  \Gamma
   \left(n{-}\frac{1}{2}\right) \Gamma \left(n {+}\frac{1}{2}\right) \Gamma
   \left(n {+} \frac{5}{2}\right)  \zeta (2 n{+}1)}{6 \pi ^{3/2} \Gamma (n{+}2)\, \lambda ^{\frac{1}{2} (2 n+1)}}\, , \\
    \mathcal{I}^{(3)}  (\lambda) &\sim \frac{\sqrt{\lambda }}{24} -\frac{23}{12} - \sum_{n=1}^{\infty} \frac{n^2  \Gamma \left(n{+}\frac{1}{2}\right)^2
   \Gamma \left(n{+}\frac{7}{2}\right) \zeta (2 n{+}1)}{6 \pi ^{3/2} \Gamma (n{+}2) \, \lambda ^{\frac{1}{2} (2 n+1)}} \, ,  \\
 \mathcal{I}^{(4)}  (\lambda) &\sim -\frac{\lambda ^{3/2}}{2880} -\frac{25 \sqrt{\lambda }}{3072} +
   2-\sum_{n=1}^{\infty} \frac{n^2 \left(40 n^3 {-}204 n^2{-}442 n {+} 27\right)  
   \Gamma \left(n{-}\frac{1}{2}\right) \Gamma \left(n{+}\frac{1}{2}\right) \Gamma
   \left(n{+}\frac{9}{2}\right)\zeta (2 n{+}1)}{11520 \pi ^{3/2} \Gamma (n{+}3)\, \lambda ^{\frac{1}{2} (2 n+1)}}  \, ,  \\
 \mathcal{I}^{(5)}  (\lambda) &\sim
 \frac{\lambda ^{3/2}}{3840} -\frac{35 \sqrt{\lambda }}{12288}  -\frac{241}{120}+ \sum_{n=1}^{\infty}\frac{n^2 (20 n^2{+}48 n{+}43)  \Gamma
   \left(n+\frac{1}{2}\right)^2 \Gamma \left(n{+}\frac{11}{2}\right) \zeta (2 n{+}1)}{11520 \pi ^{3/2} \Gamma
   (n{+}3) \, \lambda ^{\frac{1}{2} (2 n+1)} }\, .
\end{align}
It is easy to verify that these results are in agreement with \eqref{eq:applargeN1}  by considering the zero Fourier mode of the SL$(2, \mathbb{Z})$-invariant results and using the relation $\tau_2 = 4\pi N/ \lambda$.
Once more, in the above expressions we have omitted exponentially decayed terms of the form $O(e^{-\sqrt{\lambda}})$ and $O(e^{-2\sqrt{\lambda}})$ which can be derived from the large-$s$ asymptotics of the Mellin representation \eqref{eq:IMell} or more directly from the 't Hooft limit of the corresponding SL$(2, \mathbb{Z})$-invariant expressions \eqref{eq:NPDh} and $\eqref{eq:NPDf}$ respectively. 
Importantly, the leading non-perturbative corrections at order $O(N^1)$ can be immediately derived from \eqref{eq:NPDh} by computing the 't Hooft limit of the leading index $D_{N/4}(s;\tau)$ functions, whose coefficients are given in the first equation of \eqref{eq:dtilde}, and they match identically the results of \cite{Brown:2025huy}. 

\section{Giant graviton correlator at two loops  for generic $N$} \label{app:twoloops}

In this appendix we use the instrumental new results of our analysis to determine the giant graviton correlator to two loops for generic $N$. In general, the perturbative expansion of the giant graviton correlator takes the following form 
\begin{align} \label{eq:twoloop1}
 {1 \over N} \frac{  \langle  \mathcal{D}(x_1, Y_1) \mathcal{D}(x_2, Y_2) \mathcal{O}_2(x_3, Y_3) \mathcal{O}_2(x_4, Y_4) \rangle }{ \langle  \mathcal{D}(x_1, Y_1) \mathcal{D}(x_2, Y_2) \rangle} 
 = {\rm free \,\,\, part} +  \mathcal{I}_4(x_i, Y_i) \, x_{13}^4  x_{24}^4 \sum_{\ell=1}^{\infty} \left(- {\lambda \over 4\pi^2} \right)^{\ell} \sum_{\alpha}   n^{(\ell)}_{\alpha}   \, f^{(\ell)}_{\alpha}(x_i)  \, , 
\end{align}
where, compared to the expression given in equation (4) of the main text, we have introduced some additional prefactors to express the reduced correlator. The functions $f^{(\ell)}_{\alpha}(x_i)$ are $\ell$-loop conformal Feynman integrals and form a basis of integrals for the $\ell$-loop correlator, while the coefficients $n^{(\ell)}_{\alpha}$ remain to be determined.  The integrands for $f^{(\ell)}_{\alpha}(x_i)$ can be represented by the so-called $f$-graphs \cite{Eden:2012tu}; these are rational functions of $x_{ij}^2 = |x_i-x_j|^2$ with poles of the form $1/x_{ij}^2$ that have a net degree-$4$ at each point $x_i$ (this is related to the prefactor $ x_{13}^4  x_{24}^4$ we introduced so that the reduced correlator as defined in equation (4) of the main work is only a function of the cross-ratios and not the insertion points). For the correlator we study, the integrands should have an enhanced permutation symmetry $S_2 \times S_{2+\ell}$, which permutes $\{ x_1, x_2 \}$ and the integration variables $\{ x_3, x_4, \ldots, x_{\ell+4} \}$, respectively.  These properties uniquely fix the one-loop integral to be
\begin{align} \label{eq:one-loop}
    f^{(1)}(x_i)  =\int \frac{d^4 x_5}{\pi^2}   {  1   \over \prod_{1 \leq i<j \leq 5} x^2_{i,j}}\, ,   
     \end{align}
 which equals  the one-loop box  integral. At two loops there are two independent conformal integrals with the right symmetries
     \begin{align} \label{eq:two-loop}
    f^{(2)}_1(x_i)  = \int \frac{d^4 x_5}{\pi^2} \frac{d^4 x_6}{\pi^2}   {  p_1(x_i)   \over \prod_{1 \leq i<j \leq 6} x^2_{i,j}} \,  , \qquad    f^{(2)}_2(x_i)   =  \int \frac{d^4 x_5}{\pi^2} \frac{d^4 x_6}{\pi^2}   {  p_2(x_i)   \over \prod_{1 \leq i<j \leq 6} x^2_{i,j}} \,  ,
\end{align}
     where 
\begin{align} \label{eq:P1P2}
    p_1(x_i) = {1\over 16} x_{12}^2 x_{34}^2 x_{56}^2 + P_{12;3456} \, , \qquad
        p_2(x_i) = {1\over 4} x_{16}^2 x_{25}^2 x_{34}^2 + P_{12;3456} \, ,
\end{align}
and $P_{12;3456}$ denotes a  permutation on $\{1,2\}$ and $\{3,4,5,6\}$ separately.  The two-loop Feynman integrals are linear combinations of the double-box integral and the one-loop box integral squared, whose expressions are known~\cite{Usyukina:1993ch}. As shown in \cite{Wen:2022oky, Brown:2023zbr}, the integrated correlator in this representation is given by the periods of these Feynman integrals. Applying this results to our case, we derive that the integrated correlator must equal
\begin{align}
    \mathcal{C}_{\mathcal{D}} (\tau; N)\vert_{\rm pert} = -\left( {\lambda \over 4\pi^2} \right) n^{(1)}\, \mathcal{P}_{f^{(1)}} - \left( {\lambda \over 4\pi^2} \right)^2 \sum_{\alpha=1}^2 n^{(2)}_{\alpha}\, \mathcal{P}_{f^{(2)}_{\alpha}} + O(\lambda^3)\,. 
\end{align}
From~\cite{Belokurov:1983km} (see also \cite{Wen:2022oky} in the notation of the $f$-graphs), one finds that the periods of $f^{(\ell)}_\alpha$ are given by
    \begin{align} 
    \mathcal{P}_{f^{(1)}} = 6 \zeta(3) \,  , \qquad   \mathcal{P}_{f^{(2)}_1} = 3 \times 20 \zeta(5) \, , \qquad  \mathcal{P}_{f^{(2)}_2} = 12 \times 20 \zeta(5) \, . 
    \end{align} 
Thanks to our results for the integrated correlator, in particular equation (12) of the main work, we deduce that the coefficient $n^{(1)}$ at one loop is then fixed necessarily to be $n^{(1)}=1$. 

At two loops, the  integrated correlator alone is not enough to fix two unknown coefficients. {However, it is possible to combine the constraint derived from the integrated correlator with that from OPE analysis (e.g. as in \cite{Jiang:2023uut}) as to determine uniquely the two-loop correlator (essentially, two constraints fix two unknowns).} A straightforward way to obtain the two-loop result is to note that  the anomalous dimension of the Konishi operator, which dominates in the $s$-channel OPE {(i.e. the LL and HH channel)}, only receives planar contributions at the orders here considered~\cite{Fiamberti:2007rj}. This implies that any term, which is suppressed in the large-$N$ limit must be proportional to $f_1^{(2)}(x_i)$;   {this is a consequence of the fact that $f_1^{(2)}(x_i)$ has a factor of $x_{12}^2$  in the numerator (see the expression $p_1(x_i)$ in \eqref{eq:P1P2}), and is therefore 
suppressed in the OPE limit $x_{12}^2 \to 0$.} This consideration leads to the following expression for the correlator at two loops as in \eqref{eq:twoloop1},
\begin{align} \label{eq:apptwoloop1}
  -  \frac{1}{8} \left( {\lambda \over 4\pi^2} \right)^2    \left[ \left( f^{(2)}_{2}(x_i) -f^{(2)}_{1}(x_i) \right) +   c(N) \, f^{(2)}_{1}(x_i)\right] \, ,
\end{align}
where we have singled out the planar contribution (i.e. the term that is not proportional to $c(N)$) derived in \cite{Jiang:2019xdz, Jiang:2023uut}. The remaining coefficient $c(N)$, which is suppressed in large-$N$ limit, can then be fixed using the result of the integrated correlator,
\begin{align} \label{eq:apptwo-loops-corr}
    c(N) = \frac{3 N (N{+}1) {+} 2}{ N^2 (N+1)}+  \frac{ (N^2{-}1) (N{+}2)}{N^2(1 {-} (-N)^{N+1})} \, .
\end{align}
We see that indeed $c(N)$ contains the non-planar and exponentially suppressed color factors. Combining with the one-loop result, this leads to the expression given in the main text in equation (17).

Furthermore, as an important check of our results we see that for $N=2$ and $N=3$  the determinant  operator reduces to the simpler operators $\mathcal{O}_2$ and $\mathcal{O}_3$, respectively. In these cases, we find from \eqref{eq:apptwo-loops-corr} that  $c(2)=2$ while $   c(3)=1$, 
and correspondingly the two-loop correlators \eqref{eq:twoloop1} simplify and indeed match identically with the known results for the SU$(2)$ correlator $  \langle \mathcal{O}_2 
\mathcal{O}_2  \mathcal{O}_2   \mathcal{O}_2  \rangle$, and the SU$(3)$ correlator   $\langle \mathcal{O}_3 
\mathcal{O}_3  \mathcal{O}_2  \mathcal{O}_2   \rangle$, respectively. The two-loop results for these particular SU$(2)$ and SU$(3)$ correlators can be obtained directly from the corresponding planar limit expressions \cite{Eden:2012tu, Chicherin:2015edu} since these correlators do not receive non-planar corrections to the order here considered. 

Lastly, for the U$(N)$ theory we find it convenient to normalise the four-point correlator in a slightly differently manner where rather than  introducing an overall factor $1/N$ as in \eqref{eq:twoloop1}, we normalise the correlator by $N/(N^2-1)$. We then find that the correlator up to two loops is again given by exactly the same form \eqref{eq:apptwoloop1}, where we simply need to replace the coefficient $c(N)$ in \eqref{eq:apptwoloop1} with
\begin{align} \label{eq:two-loops-corrUN}
  \tilde{c}(N) = \frac{3N-2}{N(N-1)} \, .
\end{align}
Note that unlike what we saw for the SU$(N)$ case, the U$(N)$ determinant  operator for $N=2, 3$  does not reduce to $\mathcal{O}_2$ and $\mathcal{O}_3$, since in the U$(N)$ theory we have to include the operator $\mathcal{O}_1$. Nevertheless, from \eqref{eq:two-loops-corrUN} we find again that $\tilde{c}(2)=2$ thus implying that the correlator reduces to  $ \langle \mathcal{O}_2 
\mathcal{O}_2  \mathcal{O}_2   \mathcal{O}_2 \rangle$ for the U$(2)$ as well. This may be related to the fact that  $(\mathcal{O}_1)^2$ does not contribute the correlator, in which case we may identify $\cD$ with $\cO_2$ also for U$(2)$.

\vspace{0.3cm}

\twocolumngrid

\bibliography{paperFinal.bbl}

\begin{thebibliography}{77}%
\makeatletter
\providecommand \@ifxundefined [1]{%
 \@ifx{#1\undefined}
}%
\providecommand \@ifnum [1]{%
 \ifnum #1\expandafter \@firstoftwo
 \else \expandafter \@secondoftwo
 \fi
}%
\providecommand \@ifx [1]{%
 \ifx #1\expandafter \@firstoftwo
 \else \expandafter \@secondoftwo
 \fi
}%
\providecommand \natexlab [1]{#1}%
\providecommand \enquote  [1]{``#1''}%
\providecommand \bibnamefont  [1]{#1}%
\providecommand \bibfnamefont [1]{#1}%
\providecommand \citenamefont [1]{#1}%
\providecommand \href@noop [0]{\@secondoftwo}%
\providecommand \href [0]{\begingroup \@sanitize@url \@href}%
\providecommand \@href[1]{\@@startlink{#1}\@@href}%
\providecommand \@@href[1]{\endgroup#1\@@endlink}%
\providecommand \@sanitize@url [0]{\catcode `\\12\catcode `\$12\catcode
  `\&12\catcode `\#12\catcode `\^12\catcode `\_12\catcode `\%12\relax}%
\providecommand \@@startlink[1]{}%
\providecommand \@@endlink[0]{}%
\providecommand \url  [0]{\begingroup\@sanitize@url \@url }%
\providecommand \@url [1]{\endgroup\@href {#1}{\urlprefix }}%
\providecommand \urlprefix  [0]{URL }%
\providecommand \Eprint [0]{\href }%
\providecommand \doibase [0]{http://dx.doi.org/}%
\providecommand \selectlanguage [0]{\@gobble}%
\providecommand \bibinfo  [0]{\@secondoftwo}%
\providecommand \bibfield  [0]{\@secondoftwo}%
\providecommand \translation [1]{[#1]}%
\providecommand \BibitemOpen [0]{}%
\providecommand \bibitemStop [0]{}%
\providecommand \bibitemNoStop [0]{.\EOS\space}%
\providecommand \EOS [0]{\spacefactor3000\relax}%
\providecommand \BibitemShut  [1]{\csname bibitem#1\endcsname}%
\let\auto@bib@innerbib\@empty
%</preamble>
\bibitem [{\citenamefont {Freedman}\ \emph {et~al.}(1999)\citenamefont
  {Freedman}, \citenamefont {Mathur}, \citenamefont {Matusis},\ and\
  \citenamefont {Rastelli}}]{Freedman:1998tz}%
  \BibitemOpen
  \bibfield  {author} {\bibinfo {author} {\bibfnamefont {Daniel~Z.}\
  \bibnamefont {Freedman}}, \bibinfo {author} {\bibfnamefont {Samir~D.}\
  \bibnamefont {Mathur}}, \bibinfo {author} {\bibfnamefont {Alec}\ \bibnamefont
  {Matusis}}, \ and\ \bibinfo {author} {\bibfnamefont {Leonardo}\ \bibnamefont
  {Rastelli}},\ }\bibfield  {title} {\enquote {\bibinfo {title} {{Correlation
  functions in the CFT(d) / AdS(d+1) correspondence}},}\ }\href {\doibase
  10.1016/S0550-3213(99)00053-X} {\bibfield  {journal} {\bibinfo  {journal}
  {Nucl. Phys. B}\ }\textbf {\bibinfo {volume} {546}},\ \bibinfo {pages}
  {96--118} (\bibinfo {year} {1999})},\ \Eprint
  {http://arxiv.org/abs/hep-th/9804058} {arXiv:hep-th/9804058} \BibitemShut
  {NoStop}%
\bibitem [{\citenamefont {Rastelli}\ and\ \citenamefont
  {Zhou}(2017)}]{Rastelli:2016nze}%
  \BibitemOpen
  \bibfield  {author} {\bibinfo {author} {\bibfnamefont {Leonardo}\
  \bibnamefont {Rastelli}}\ and\ \bibinfo {author} {\bibfnamefont {Xinan}\
  \bibnamefont {Zhou}},\ }\bibfield  {title} {\enquote {\bibinfo {title}
  {{Mellin amplitudes for $AdS_5\times S^5$}},}\ }\href {\doibase
  10.1103/PhysRevLett.118.091602} {\bibfield  {journal} {\bibinfo  {journal}
  {Phys. Rev. Lett.}\ }\textbf {\bibinfo {volume} {118}},\ \bibinfo {pages}
  {091602} (\bibinfo {year} {2017})},\ \Eprint
  {http://arxiv.org/abs/1608.06624} {arXiv:1608.06624 [hep-th]} \BibitemShut
  {NoStop}%
\bibitem [{\citenamefont {Aprile}\ \emph {et~al.}(2018)\citenamefont {Aprile},
  \citenamefont {Drummond}, \citenamefont {Heslop},\ and\ \citenamefont
  {Paul}}]{Aprile:2017bgs}%
  \BibitemOpen
  \bibfield  {author} {\bibinfo {author} {\bibfnamefont {F.}~\bibnamefont
  {Aprile}}, \bibinfo {author} {\bibfnamefont {J.~M.}\ \bibnamefont
  {Drummond}}, \bibinfo {author} {\bibfnamefont {P.}~\bibnamefont {Heslop}}, \
  and\ \bibinfo {author} {\bibfnamefont {H.}~\bibnamefont {Paul}},\ }\bibfield
  {title} {\enquote {\bibinfo {title} {{Quantum Gravity from Conformal Field
  Theory}},}\ }\href {\doibase 10.1007/JHEP01(2018)035} {\bibfield  {journal}
  {\bibinfo  {journal} {JHEP}\ }\textbf {\bibinfo {volume} {01}},\ \bibinfo
  {pages} {035} (\bibinfo {year} {2018})},\ \Eprint
  {http://arxiv.org/abs/1706.02822} {arXiv:1706.02822 [hep-th]} \BibitemShut
  {NoStop}%
\bibitem [{\citenamefont {Caron-Huot}\ and\ \citenamefont
  {Trinh}(2019)}]{Caron-Huot:2018kta}%
  \BibitemOpen
  \bibfield  {author} {\bibinfo {author} {\bibfnamefont {Simon}\ \bibnamefont
  {Caron-Huot}}\ and\ \bibinfo {author} {\bibfnamefont {Anh-Khoi}\ \bibnamefont
  {Trinh}},\ }\bibfield  {title} {\enquote {\bibinfo {title} {{All tree-level
  correlators in AdS$_{5}${\texttimes}S$_{5}$ supergravity: hidden
  ten-dimensional conformal symmetry}},}\ }\href {\doibase
  10.1007/JHEP01(2019)196} {\bibfield  {journal} {\bibinfo  {journal} {JHEP}\
  }\textbf {\bibinfo {volume} {01}},\ \bibinfo {pages} {196} (\bibinfo {year}
  {2019})},\ \Eprint {http://arxiv.org/abs/1809.09173} {arXiv:1809.09173
  [hep-th]} \BibitemShut {NoStop}%
\bibitem [{\citenamefont {Alday}\ and\ \citenamefont
  {Zhou}(2020)}]{Alday:2019nin}%
  \BibitemOpen
  \bibfield  {author} {\bibinfo {author} {\bibfnamefont {Luis~F.}\ \bibnamefont
  {Alday}}\ and\ \bibinfo {author} {\bibfnamefont {Xinan}\ \bibnamefont
  {Zhou}},\ }\bibfield  {title} {\enquote {\bibinfo {title} {{Simplicity of AdS
  Supergravity at One Loop}},}\ }\href {\doibase 10.1007/JHEP09(2020)008}
  {\bibfield  {journal} {\bibinfo  {journal} {JHEP}\ }\textbf {\bibinfo
  {volume} {09}},\ \bibinfo {pages} {008} (\bibinfo {year} {2020})},\ \Eprint
  {http://arxiv.org/abs/1912.02663} {arXiv:1912.02663 [hep-th]} \BibitemShut
  {NoStop}%
\bibitem [{\citenamefont {Alday}\ and\ \citenamefont
  {Hansen}(2023)}]{Alday:2023mvu}%
  \BibitemOpen
  \bibfield  {author} {\bibinfo {author} {\bibfnamefont {Luis~F.}\ \bibnamefont
  {Alday}}\ and\ \bibinfo {author} {\bibfnamefont {Tobias}\ \bibnamefont
  {Hansen}},\ }\bibfield  {title} {\enquote {\bibinfo {title} {{The AdS
  Virasoro-Shapiro amplitude}},}\ }\href {\doibase 10.1007/JHEP10(2023)023}
  {\bibfield  {journal} {\bibinfo  {journal} {JHEP}\ }\textbf {\bibinfo
  {volume} {10}},\ \bibinfo {pages} {023} (\bibinfo {year} {2023})},\ \Eprint
  {http://arxiv.org/abs/2306.12786} {arXiv:2306.12786 [hep-th]} \BibitemShut
  {NoStop}%
\bibitem [{\citenamefont {Aprile}\ \emph {et~al.}(2025)\citenamefont {Aprile},
  \citenamefont {Giusto},\ and\ \citenamefont {Russo}}]{Aprile:2024lwy}%
  \BibitemOpen
  \bibfield  {author} {\bibinfo {author} {\bibfnamefont {Francesco}\
  \bibnamefont {Aprile}}, \bibinfo {author} {\bibfnamefont {Stefano}\
  \bibnamefont {Giusto}}, \ and\ \bibinfo {author} {\bibfnamefont {Rodolfo}\
  \bibnamefont {Russo}},\ }\bibfield  {title} {\enquote {\bibinfo {title}
  {{Holographic correlators with BPS bound states in $\mathcal{N} = 4$ SYM}},}\
  }\href {\doibase 10.1103/PhysRevLett.134.091602} {\bibfield  {journal}
  {\bibinfo  {journal} {Phys. Rev. Lett.}\ }\textbf {\bibinfo {volume} {134}},\
  \bibinfo {pages} {091602} (\bibinfo {year} {2025})},\ \Eprint
  {http://arxiv.org/abs/2409.12911} {arXiv:2409.12911 [hep-th]} \BibitemShut
  {NoStop}%
\bibitem [{\citenamefont {Wang}\ \emph {et~al.}(2025)\citenamefont {Wang},
  \citenamefont {Wu},\ and\ \citenamefont {Yuan}}]{Wang:2025pjo}%
  \BibitemOpen
  \bibfield  {author} {\bibinfo {author} {\bibfnamefont {Bo}~\bibnamefont
  {Wang}}, \bibinfo {author} {\bibfnamefont {Di}~\bibnamefont {Wu}}, \ and\
  \bibinfo {author} {\bibfnamefont {Ellis~Ye}\ \bibnamefont {Yuan}},\
  }\bibfield  {title} {\enquote {\bibinfo {title} {{Kaluza-Klein AdS
  Virasoro-Shapiro Amplitude near Flat Space}},}\ }\href {\doibase
  10.1103/v72s-rv7y} {\bibfield  {journal} {\bibinfo  {journal} {Phys. Rev.
  Lett.}\ }\textbf {\bibinfo {volume} {135}},\ \bibinfo {pages} {041603}
  (\bibinfo {year} {2025})},\ \Eprint {http://arxiv.org/abs/2503.01964}
  {arXiv:2503.01964 [hep-th]} \BibitemShut {NoStop}%
\bibitem [{\citenamefont {Heslop}(2022)}]{Heslop:2022xgp}%
  \BibitemOpen
  \bibfield  {author} {\bibinfo {author} {\bibfnamefont {Paul}\ \bibnamefont
  {Heslop}},\ }\bibfield  {title} {\enquote {\bibinfo {title} {{The SAGEX
  Review on Scattering Amplitudes, Chapter 8: Half BPS correlators}},}\ }\href
  {\doibase 10.1088/1751-8121/ac8c71} {\bibfield  {journal} {\bibinfo
  {journal} {J. Phys. A}\ }\textbf {\bibinfo {volume} {55}},\ \bibinfo {pages}
  {443009} (\bibinfo {year} {2022})},\ \Eprint
  {http://arxiv.org/abs/2203.13019} {arXiv:2203.13019 [hep-th]} \BibitemShut
  {NoStop}%
\bibitem [{\citenamefont {Bissi}\ \emph {et~al.}(2022)\citenamefont {Bissi},
  \citenamefont {Sinha},\ and\ \citenamefont {Zhou}}]{Bissi:2022mrs}%
  \BibitemOpen
  \bibfield  {author} {\bibinfo {author} {\bibfnamefont {Agnese}\ \bibnamefont
  {Bissi}}, \bibinfo {author} {\bibfnamefont {Aninda}\ \bibnamefont {Sinha}}, \
  and\ \bibinfo {author} {\bibfnamefont {Xinan}\ \bibnamefont {Zhou}},\
  }\bibfield  {title} {\enquote {\bibinfo {title} {{Selected topics in analytic
  conformal bootstrap: A guided journey}},}\ }\href {\doibase
  10.1016/j.physrep.2022.09.004} {\bibfield  {journal} {\bibinfo  {journal}
  {Phys. Rept.}\ }\textbf {\bibinfo {volume} {991}},\ \bibinfo {pages} {1--89}
  (\bibinfo {year} {2022})},\ \Eprint {http://arxiv.org/abs/2202.08475}
  {arXiv:2202.08475 [hep-th]} \BibitemShut {NoStop}%
\bibitem [{\citenamefont {Binder}\ \emph {et~al.}(2019)\citenamefont {Binder},
  \citenamefont {Chester}, \citenamefont {Pufu},\ and\ \citenamefont
  {Wang}}]{Binder:2019jwn}%
  \BibitemOpen
  \bibfield  {author} {\bibinfo {author} {\bibfnamefont {Damon~J.}\
  \bibnamefont {Binder}}, \bibinfo {author} {\bibfnamefont {Shai~M.}\
  \bibnamefont {Chester}}, \bibinfo {author} {\bibfnamefont {Silviu~S.}\
  \bibnamefont {Pufu}}, \ and\ \bibinfo {author} {\bibfnamefont {Yifan}\
  \bibnamefont {Wang}},\ }\bibfield  {title} {\enquote {\bibinfo {title} {{$
  \mathcal{N} $ = 4 Super-Yang-Mills correlators at strong coupling from string
  theory and localization}},}\ }\href {\doibase 10.1007/JHEP12(2019)119}
  {\bibfield  {journal} {\bibinfo  {journal} {JHEP}\ }\textbf {\bibinfo
  {volume} {12}},\ \bibinfo {pages} {119} (\bibinfo {year} {2019})},\ \Eprint
  {http://arxiv.org/abs/1902.06263} {arXiv:1902.06263 [hep-th]} \BibitemShut
  {NoStop}%
\bibitem [{\citenamefont {Chester}\ and\ \citenamefont
  {Pufu}(2021)}]{Chester:2020dja}%
  \BibitemOpen
  \bibfield  {author} {\bibinfo {author} {\bibfnamefont {Shai~M.}\ \bibnamefont
  {Chester}}\ and\ \bibinfo {author} {\bibfnamefont {Silviu~S.}\ \bibnamefont
  {Pufu}},\ }\bibfield  {title} {\enquote {\bibinfo {title} {{Far beyond the
  planar limit in strongly-coupled $ \mathcal{N} $ = 4 SYM}},}\ }\href
  {\doibase 10.1007/JHEP01(2021)103} {\bibfield  {journal} {\bibinfo  {journal}
  {JHEP}\ }\textbf {\bibinfo {volume} {01}},\ \bibinfo {pages} {103} (\bibinfo
  {year} {2021})},\ \Eprint {http://arxiv.org/abs/2003.08412} {arXiv:2003.08412
  [hep-th]} \BibitemShut {NoStop}%
\bibitem [{\citenamefont {Dorigoni}\ \emph
  {et~al.}(2022{\natexlab{a}})\citenamefont {Dorigoni}, \citenamefont {Green},\
  and\ \citenamefont {Wen}}]{Dorigoni:2022iem}%
  \BibitemOpen
  \bibfield  {author} {\bibinfo {author} {\bibfnamefont {Daniele}\ \bibnamefont
  {Dorigoni}}, \bibinfo {author} {\bibfnamefont {Michael~B.}\ \bibnamefont
  {Green}}, \ and\ \bibinfo {author} {\bibfnamefont {Congkao}\ \bibnamefont
  {Wen}},\ }\bibfield  {title} {\enquote {\bibinfo {title} {{The SAGEX review
  on scattering amplitudes Chapter 10: Selected topics on modular covariance of
  type IIB string amplitudes and their~~supersymmetric Yang{\textendash}Mills
  duals}},}\ }\href {\doibase 10.1088/1751-8121/ac9263} {\bibfield  {journal}
  {\bibinfo  {journal} {J. Phys. A}\ }\textbf {\bibinfo {volume} {55}},\
  \bibinfo {pages} {443011} (\bibinfo {year} {2022}{\natexlab{a}})},\ \Eprint
  {http://arxiv.org/abs/2203.13021} {arXiv:2203.13021 [hep-th]} \BibitemShut
  {NoStop}%
\bibitem [{\citenamefont {Chester}\ \emph {et~al.}(2020)\citenamefont
  {Chester}, \citenamefont {Green}, \citenamefont {Pufu}, \citenamefont
  {Wang},\ and\ \citenamefont {Wen}}]{Chester:2019jas}%
  \BibitemOpen
  \bibfield  {author} {\bibinfo {author} {\bibfnamefont {Shai~M.}\ \bibnamefont
  {Chester}}, \bibinfo {author} {\bibfnamefont {Michael~B.}\ \bibnamefont
  {Green}}, \bibinfo {author} {\bibfnamefont {Silviu~S.}\ \bibnamefont {Pufu}},
  \bibinfo {author} {\bibfnamefont {Yifan}\ \bibnamefont {Wang}}, \ and\
  \bibinfo {author} {\bibfnamefont {Congkao}\ \bibnamefont {Wen}},\ }\bibfield
  {title} {\enquote {\bibinfo {title} {{Modular invariance in superstring
  theory from $ \mathcal{N} $ = 4 super-Yang-Mills}},}\ }\href {\doibase
  10.1007/JHEP11(2020)016} {\bibfield  {journal} {\bibinfo  {journal} {JHEP}\
  }\textbf {\bibinfo {volume} {11}},\ \bibinfo {pages} {016} (\bibinfo {year}
  {2020})},\ \Eprint {http://arxiv.org/abs/1912.13365} {arXiv:1912.13365
  [hep-th]} \BibitemShut {NoStop}%
\bibitem [{\citenamefont {Chester}\ \emph {et~al.}(2021)\citenamefont
  {Chester}, \citenamefont {Green}, \citenamefont {Pufu}, \citenamefont
  {Wang},\ and\ \citenamefont {Wen}}]{Chester:2020vyz}%
  \BibitemOpen
  \bibfield  {author} {\bibinfo {author} {\bibfnamefont {Shai~M.}\ \bibnamefont
  {Chester}}, \bibinfo {author} {\bibfnamefont {Michael~B.}\ \bibnamefont
  {Green}}, \bibinfo {author} {\bibfnamefont {Silviu~S.}\ \bibnamefont {Pufu}},
  \bibinfo {author} {\bibfnamefont {Yifan}\ \bibnamefont {Wang}}, \ and\
  \bibinfo {author} {\bibfnamefont {Congkao}\ \bibnamefont {Wen}},\ }\bibfield
  {title} {\enquote {\bibinfo {title} {{New modular invariants in $ \mathcal{N}
  $ = 4 Super-Yang-Mills theory}},}\ }\href {\doibase 10.1007/JHEP04(2021)212}
  {\bibfield  {journal} {\bibinfo  {journal} {JHEP}\ }\textbf {\bibinfo
  {volume} {04}},\ \bibinfo {pages} {212} (\bibinfo {year} {2021})},\ \Eprint
  {http://arxiv.org/abs/2008.02713} {arXiv:2008.02713 [hep-th]} \BibitemShut
  {NoStop}%
\bibitem [{\citenamefont {Dorigoni}\ \emph
  {et~al.}(2021{\natexlab{a}})\citenamefont {Dorigoni}, \citenamefont {Green},\
  and\ \citenamefont {Wen}}]{Dorigoni:2021guq}%
  \BibitemOpen
  \bibfield  {author} {\bibinfo {author} {\bibfnamefont {Daniele}\ \bibnamefont
  {Dorigoni}}, \bibinfo {author} {\bibfnamefont {Michael~B.}\ \bibnamefont
  {Green}}, \ and\ \bibinfo {author} {\bibfnamefont {Congkao}\ \bibnamefont
  {Wen}},\ }\bibfield  {title} {\enquote {\bibinfo {title} {{Exact properties
  of an integrated correlator in $ \mathcal{N} $ = 4 SU(N) SYM}},}\ }\href
  {\doibase 10.1007/JHEP05(2021)089} {\bibfield  {journal} {\bibinfo  {journal}
  {JHEP}\ }\textbf {\bibinfo {volume} {05}},\ \bibinfo {pages} {089} (\bibinfo
  {year} {2021}{\natexlab{a}})},\ \Eprint {http://arxiv.org/abs/2102.09537}
  {arXiv:2102.09537 [hep-th]} \BibitemShut {NoStop}%
\bibitem [{\citenamefont {Dorigoni}\ \emph
  {et~al.}(2021{\natexlab{b}})\citenamefont {Dorigoni}, \citenamefont {Green},\
  and\ \citenamefont {Wen}}]{Dorigoni:2021bvj}%
  \BibitemOpen
  \bibfield  {author} {\bibinfo {author} {\bibfnamefont {Daniele}\ \bibnamefont
  {Dorigoni}}, \bibinfo {author} {\bibfnamefont {Michael~B.}\ \bibnamefont
  {Green}}, \ and\ \bibinfo {author} {\bibfnamefont {Congkao}\ \bibnamefont
  {Wen}},\ }\bibfield  {title} {\enquote {\bibinfo {title} {{Novel
  Representation of an Integrated Correlator in $\mathcal N$ = 4 Supersymmetric
  Yang-Mills Theory}},}\ }\href {\doibase 10.1103/PhysRevLett.126.161601}
  {\bibfield  {journal} {\bibinfo  {journal} {Phys. Rev. Lett.}\ }\textbf
  {\bibinfo {volume} {126}},\ \bibinfo {pages} {161601} (\bibinfo {year}
  {2021}{\natexlab{b}})},\ \Eprint {http://arxiv.org/abs/2102.08305}
  {arXiv:2102.08305 [hep-th]} \BibitemShut {NoStop}%
\bibitem [{\citenamefont {Alday}\ \emph {et~al.}(2024)\citenamefont {Alday},
  \citenamefont {Chester}, \citenamefont {Dorigoni}, \citenamefont {Green},\
  and\ \citenamefont {Wen}}]{Alday:2023pet}%
  \BibitemOpen
  \bibfield  {author} {\bibinfo {author} {\bibfnamefont {Luis~F.}\ \bibnamefont
  {Alday}}, \bibinfo {author} {\bibfnamefont {Shai~M.}\ \bibnamefont
  {Chester}}, \bibinfo {author} {\bibfnamefont {Daniele}\ \bibnamefont
  {Dorigoni}}, \bibinfo {author} {\bibfnamefont {Michael~B.}\ \bibnamefont
  {Green}}, \ and\ \bibinfo {author} {\bibfnamefont {Congkao}\ \bibnamefont
  {Wen}},\ }\bibfield  {title} {\enquote {\bibinfo {title} {{Relations between
  integrated correlators in $ \mathcal{N} $ = 4 supersymmetric Yang-Mills
  theory}},}\ }\href {\doibase 10.1007/JHEP05(2024)044} {\bibfield  {journal}
  {\bibinfo  {journal} {JHEP}\ }\textbf {\bibinfo {volume} {05}},\ \bibinfo
  {pages} {044} (\bibinfo {year} {2024})},\ \Eprint
  {http://arxiv.org/abs/2310.12322} {arXiv:2310.12322 [hep-th]} \BibitemShut
  {NoStop}%
\bibitem [{\citenamefont {Chester}\ \emph {et~al.}(2023)\citenamefont
  {Chester}, \citenamefont {Dempsey},\ and\ \citenamefont
  {Pufu}}]{Chester:2021aun}%
  \BibitemOpen
  \bibfield  {author} {\bibinfo {author} {\bibfnamefont {Shai~M.}\ \bibnamefont
  {Chester}}, \bibinfo {author} {\bibfnamefont {Ross}\ \bibnamefont {Dempsey}},
  \ and\ \bibinfo {author} {\bibfnamefont {Silviu~S.}\ \bibnamefont {Pufu}},\
  }\bibfield  {title} {\enquote {\bibinfo {title} {{Bootstrapping $ \mathcal{N}
  $ = 4 super-Yang-Mills on the conformal manifold}},}\ }\href {\doibase
  10.1007/JHEP01(2023)038} {\bibfield  {journal} {\bibinfo  {journal} {JHEP}\
  }\textbf {\bibinfo {volume} {01}},\ \bibinfo {pages} {038} (\bibinfo {year}
  {2023})},\ \Eprint {http://arxiv.org/abs/2111.07989} {arXiv:2111.07989
  [hep-th]} \BibitemShut {NoStop}%
\bibitem [{\citenamefont {Caron-Huot}\ \emph {et~al.}(2025)\citenamefont
  {Caron-Huot}, \citenamefont {Coronado},\ and\ \citenamefont
  {Zahraee}}]{Caron-Huot:2024tzr}%
  \BibitemOpen
  \bibfield  {author} {\bibinfo {author} {\bibfnamefont {Simon}\ \bibnamefont
  {Caron-Huot}}, \bibinfo {author} {\bibfnamefont {Frank}\ \bibnamefont
  {Coronado}}, \ and\ \bibinfo {author} {\bibfnamefont {Zahra}\ \bibnamefont
  {Zahraee}},\ }\bibfield  {title} {\enquote {\bibinfo {title} {{Bootstrapping
  $ \mathcal{N} $ = 4 sYM correlators using integrability and localization}},}\
  }\href {\doibase 10.1007/JHEP05(2025)220} {\bibfield  {journal} {\bibinfo
  {journal} {JHEP}\ }\textbf {\bibinfo {volume} {05}},\ \bibinfo {pages} {220}
  (\bibinfo {year} {2025})},\ \Eprint {http://arxiv.org/abs/2412.00249}
  {arXiv:2412.00249 [hep-th]} \BibitemShut {NoStop}%
\bibitem [{\citenamefont {Dempsey}\ \emph {et~al.}(2025)\citenamefont
  {Dempsey}, \citenamefont {Karlsson}, \citenamefont {Pufu}, \citenamefont
  {Zahraee},\ and\ \citenamefont {Zhiboedov}}]{Dempsey:2025yiv}%
  \BibitemOpen
  \bibfield  {author} {\bibinfo {author} {\bibfnamefont {Ross}\ \bibnamefont
  {Dempsey}}, \bibinfo {author} {\bibfnamefont {Robin}\ \bibnamefont
  {Karlsson}}, \bibinfo {author} {\bibfnamefont {Silviu~S.}\ \bibnamefont
  {Pufu}}, \bibinfo {author} {\bibfnamefont {Zahra}\ \bibnamefont {Zahraee}}, \
  and\ \bibinfo {author} {\bibfnamefont {Alexander}\ \bibnamefont
  {Zhiboedov}},\ }\bibfield  {title} {\enquote {\bibinfo {title} {{Conformal
  collider bootstrap in ${\mathcal N}=4$ SYM}},}\ }\href@noop {} {\  (\bibinfo
  {year} {2025})},\ \Eprint {http://arxiv.org/abs/2512.10796} {arXiv:2512.10796
  [hep-th]} \BibitemShut {NoStop}%
\bibitem [{\citenamefont {McGreevy}\ \emph {et~al.}(2000)\citenamefont
  {McGreevy}, \citenamefont {Susskind},\ and\ \citenamefont
  {Toumbas}}]{McGreevy:2000cw}%
  \BibitemOpen
  \bibfield  {author} {\bibinfo {author} {\bibfnamefont {John}\ \bibnamefont
  {McGreevy}}, \bibinfo {author} {\bibfnamefont {Leonard}\ \bibnamefont
  {Susskind}}, \ and\ \bibinfo {author} {\bibfnamefont {Nicolaos}\ \bibnamefont
  {Toumbas}},\ }\bibfield  {title} {\enquote {\bibinfo {title} {{Invasion of
  the giant gravitons from Anti-de Sitter space}},}\ }\href {\doibase
  10.1088/1126-6708/2000/06/008} {\bibfield  {journal} {\bibinfo  {journal}
  {JHEP}\ }\textbf {\bibinfo {volume} {06}},\ \bibinfo {pages} {008} (\bibinfo
  {year} {2000})},\ \Eprint {http://arxiv.org/abs/hep-th/0003075}
  {arXiv:hep-th/0003075} \BibitemShut {NoStop}%
\bibitem [{\citenamefont {Hashimoto}\ \emph {et~al.}(2000)\citenamefont
  {Hashimoto}, \citenamefont {Hirano},\ and\ \citenamefont
  {Itzhaki}}]{Hashimoto:2000zp}%
  \BibitemOpen
  \bibfield  {author} {\bibinfo {author} {\bibfnamefont {Akikazu}\ \bibnamefont
  {Hashimoto}}, \bibinfo {author} {\bibfnamefont {Shinji}\ \bibnamefont
  {Hirano}}, \ and\ \bibinfo {author} {\bibfnamefont {N.}~\bibnamefont
  {Itzhaki}},\ }\bibfield  {title} {\enquote {\bibinfo {title} {{Large branes
  in AdS and their field theory dual}},}\ }\href {\doibase
  10.1088/1126-6708/2000/08/051} {\bibfield  {journal} {\bibinfo  {journal}
  {JHEP}\ }\textbf {\bibinfo {volume} {08}},\ \bibinfo {pages} {051} (\bibinfo
  {year} {2000})},\ \Eprint {http://arxiv.org/abs/hep-th/0008016}
  {arXiv:hep-th/0008016} \BibitemShut {NoStop}%
\bibitem [{\citenamefont {Balasubramanian}\ \emph {et~al.}(2002)\citenamefont
  {Balasubramanian}, \citenamefont {Berkooz}, \citenamefont {Naqvi},\ and\
  \citenamefont {Strassler}}]{Balasubramanian:2001nh}%
  \BibitemOpen
  \bibfield  {author} {\bibinfo {author} {\bibfnamefont {Vijay}\ \bibnamefont
  {Balasubramanian}}, \bibinfo {author} {\bibfnamefont {Micha}\ \bibnamefont
  {Berkooz}}, \bibinfo {author} {\bibfnamefont {Asad}\ \bibnamefont {Naqvi}}, \
  and\ \bibinfo {author} {\bibfnamefont {Matthew~J.}\ \bibnamefont
  {Strassler}},\ }\bibfield  {title} {\enquote {\bibinfo {title} {{Giant
  gravitons in conformal field theory}},}\ }\href {\doibase
  10.1088/1126-6708/2002/04/034} {\bibfield  {journal} {\bibinfo  {journal}
  {JHEP}\ }\textbf {\bibinfo {volume} {04}},\ \bibinfo {pages} {034} (\bibinfo
  {year} {2002})},\ \Eprint {http://arxiv.org/abs/hep-th/0107119}
  {arXiv:hep-th/0107119} \BibitemShut {NoStop}%
\bibitem [{\citenamefont {Corley}\ \emph {et~al.}(2002)\citenamefont {Corley},
  \citenamefont {Jevicki},\ and\ \citenamefont {Ramgoolam}}]{Corley:2001zk}%
  \BibitemOpen
  \bibfield  {author} {\bibinfo {author} {\bibfnamefont {Steve}\ \bibnamefont
  {Corley}}, \bibinfo {author} {\bibfnamefont {Antal}\ \bibnamefont {Jevicki}},
  \ and\ \bibinfo {author} {\bibfnamefont {Sanjaye}\ \bibnamefont
  {Ramgoolam}},\ }\bibfield  {title} {\enquote {\bibinfo {title} {{Exact
  correlators of giant gravitons from dual N=4 SYM theory}},}\ }\href {\doibase
  10.4310/ATMP.2001.v5.n4.a6} {\bibfield  {journal} {\bibinfo  {journal} {Adv.
  Theor. Math. Phys.}\ }\textbf {\bibinfo {volume} {5}},\ \bibinfo {pages}
  {809--839} (\bibinfo {year} {2002})},\ \Eprint
  {http://arxiv.org/abs/hep-th/0111222} {arXiv:hep-th/0111222} \BibitemShut
  {NoStop}%
\bibitem [{\citenamefont {Brown}\ \emph {et~al.}(2025)\citenamefont {Brown},
  \citenamefont {Dorigoni}, \citenamefont {Galvagno},\ and\ \citenamefont
  {Wen}}]{Brown:2025huy}%
  \BibitemOpen
  \bibfield  {author} {\bibinfo {author} {\bibfnamefont {Augustus}\
  \bibnamefont {Brown}}, \bibinfo {author} {\bibfnamefont {Daniele}\
  \bibnamefont {Dorigoni}}, \bibinfo {author} {\bibfnamefont {Francesco}\
  \bibnamefont {Galvagno}}, \ and\ \bibinfo {author} {\bibfnamefont {Congkao}\
  \bibnamefont {Wen}},\ }\bibfield  {title} {\enquote {\bibinfo {title}
  {{Universality of giant graviton correlators}},}\ }\href {\doibase
  10.1007/JHEP11(2025)034} {\bibfield  {journal} {\bibinfo  {journal} {JHEP}\
  }\textbf {\bibinfo {volume} {11}},\ \bibinfo {pages} {034} (\bibinfo {year}
  {2025})},\ \Eprint {http://arxiv.org/abs/2508.15657} {arXiv:2508.15657
  [hep-th]} \BibitemShut {NoStop}%
\bibitem [{\citenamefont {Imamura}(2021)}]{Imamura:2021ytr}%
  \BibitemOpen
  \bibfield  {author} {\bibinfo {author} {\bibfnamefont {Yosuke}\ \bibnamefont
  {Imamura}},\ }\bibfield  {title} {\enquote {\bibinfo {title} {{Finite-N
  superconformal index via the AdS/CFT correspondence}},}\ }\href {\doibase
  10.1093/ptep/ptab141} {\bibfield  {journal} {\bibinfo  {journal} {PTEP}\
  }\textbf {\bibinfo {volume} {2021}},\ \bibinfo {pages} {123B05} (\bibinfo
  {year} {2021})},\ \Eprint {http://arxiv.org/abs/2108.12090} {arXiv:2108.12090
  [hep-th]} \BibitemShut {NoStop}%
\bibitem [{\citenamefont {Gaiotto}\ and\ \citenamefont
  {Lee}(2024)}]{Gaiotto:2021xce}%
  \BibitemOpen
  \bibfield  {author} {\bibinfo {author} {\bibfnamefont {Davide}\ \bibnamefont
  {Gaiotto}}\ and\ \bibinfo {author} {\bibfnamefont {Ji~Hoon}\ \bibnamefont
  {Lee}},\ }\bibfield  {title} {\enquote {\bibinfo {title} {{The giant graviton
  expansion}},}\ }\href {\doibase 10.1007/JHEP08(2024)025} {\bibfield
  {journal} {\bibinfo  {journal} {JHEP}\ }\textbf {\bibinfo {volume} {08}},\
  \bibinfo {pages} {025} (\bibinfo {year} {2024})},\ \Eprint
  {http://arxiv.org/abs/2109.02545} {arXiv:2109.02545 [hep-th]} \BibitemShut
  {NoStop}%
\bibitem [{\citenamefont {Eleftheriou}\ \emph {et~al.}(2024)\citenamefont
  {Eleftheriou}, \citenamefont {Murthy},\ and\ \citenamefont
  {Rossell{\'o}}}]{Eleftheriou:2023jxr}%
  \BibitemOpen
  \bibfield  {author} {\bibinfo {author} {\bibfnamefont {Giorgos}\ \bibnamefont
  {Eleftheriou}}, \bibinfo {author} {\bibfnamefont {Sameer}\ \bibnamefont
  {Murthy}}, \ and\ \bibinfo {author} {\bibfnamefont {Mart{\'\i}}\ \bibnamefont
  {Rossell{\'o}}},\ }\bibfield  {title} {\enquote {\bibinfo {title} {{The giant
  graviton expansion in $AdS_5 \times S^5$}},}\ }\href {\doibase
  10.21468/SciPostPhys.17.4.098} {\bibfield  {journal} {\bibinfo  {journal}
  {SciPost Phys.}\ }\textbf {\bibinfo {volume} {17}},\ \bibinfo {pages} {098}
  (\bibinfo {year} {2024})},\ \Eprint {http://arxiv.org/abs/2312.14921}
  {arXiv:2312.14921 [hep-th]} \BibitemShut {NoStop}%
\bibitem [{\citenamefont {Witten}(1979)}]{Witten:1979kh}%
  \BibitemOpen
  \bibfield  {author} {\bibinfo {author} {\bibfnamefont {Edward}\ \bibnamefont
  {Witten}},\ }\bibfield  {title} {\enquote {\bibinfo {title} {{Baryons in the
  1/n Expansion}},}\ }\href {\doibase 10.1016/0550-3213(79)90232-3} {\bibfield
  {journal} {\bibinfo  {journal} {Nucl. Phys. B}\ }\textbf {\bibinfo {volume}
  {160}},\ \bibinfo {pages} {57--115} (\bibinfo {year} {1979})}\BibitemShut
  {NoStop}%
\bibitem [{\citenamefont {Witten}(1998)}]{Witten:1998xy}%
  \BibitemOpen
  \bibfield  {author} {\bibinfo {author} {\bibfnamefont {Edward}\ \bibnamefont
  {Witten}},\ }\bibfield  {title} {\enquote {\bibinfo {title} {{Baryons and
  branes in anti-de Sitter space}},}\ }\href {\doibase
  10.1088/1126-6708/1998/07/006} {\bibfield  {journal} {\bibinfo  {journal}
  {JHEP}\ }\textbf {\bibinfo {volume} {07}},\ \bibinfo {pages} {006} (\bibinfo
  {year} {1998})},\ \Eprint {http://arxiv.org/abs/hep-th/9805112}
  {arXiv:hep-th/9805112} \BibitemShut {NoStop}%
\bibitem [{\citenamefont {de~Mello~Koch}\ and\ \citenamefont
  {Gwyn}(2004)}]{deMelloKoch:2004crq}%
  \BibitemOpen
  \bibfield  {author} {\bibinfo {author} {\bibfnamefont {Robert}\ \bibnamefont
  {de~Mello~Koch}}\ and\ \bibinfo {author} {\bibfnamefont {Rhiannon}\
  \bibnamefont {Gwyn}},\ }\bibfield  {title} {\enquote {\bibinfo {title}
  {{Giant graviton correlators from dual SU(N) super Yang-Mills theory}},}\
  }\href {\doibase 10.1088/1126-6708/2004/11/081} {\bibfield  {journal}
  {\bibinfo  {journal} {JHEP}\ }\textbf {\bibinfo {volume} {11}},\ \bibinfo
  {pages} {081} (\bibinfo {year} {2004})},\ \Eprint
  {http://arxiv.org/abs/hep-th/0410236} {arXiv:hep-th/0410236} \BibitemShut
  {NoStop}%
\bibitem [{\citenamefont {Kimura}\ and\ \citenamefont
  {Ramgoolam}(2007)}]{Kimura:2007wy}%
  \BibitemOpen
  \bibfield  {author} {\bibinfo {author} {\bibfnamefont {Yusuke}\ \bibnamefont
  {Kimura}}\ and\ \bibinfo {author} {\bibfnamefont {Sanjaye}\ \bibnamefont
  {Ramgoolam}},\ }\bibfield  {title} {\enquote {\bibinfo {title} {{Branes,
  anti-branes and brauer algebras in gauge-gravity duality}},}\ }\href
  {\doibase 10.1088/1126-6708/2007/11/078} {\bibfield  {journal} {\bibinfo
  {journal} {JHEP}\ }\textbf {\bibinfo {volume} {11}},\ \bibinfo {pages} {078}
  (\bibinfo {year} {2007})},\ \Eprint {http://arxiv.org/abs/0709.2158}
  {arXiv:0709.2158 [hep-th]} \BibitemShut {NoStop}%
\bibitem [{\citenamefont {Bissi}\ \emph {et~al.}(2011)\citenamefont {Bissi},
  \citenamefont {Kristjansen}, \citenamefont {Young},\ and\ \citenamefont
  {Zoubos}}]{Bissi:2011dc}%
  \BibitemOpen
  \bibfield  {author} {\bibinfo {author} {\bibfnamefont {A.}~\bibnamefont
  {Bissi}}, \bibinfo {author} {\bibfnamefont {C.}~\bibnamefont {Kristjansen}},
  \bibinfo {author} {\bibfnamefont {D.}~\bibnamefont {Young}}, \ and\ \bibinfo
  {author} {\bibfnamefont {K.}~\bibnamefont {Zoubos}},\ }\bibfield  {title}
  {\enquote {\bibinfo {title} {{Holographic three-point functions of giant
  gravitons}},}\ }\href {\doibase 10.1007/JHEP06(2011)085} {\bibfield
  {journal} {\bibinfo  {journal} {JHEP}\ }\textbf {\bibinfo {volume} {06}},\
  \bibinfo {pages} {085} (\bibinfo {year} {2011})},\ \Eprint
  {http://arxiv.org/abs/1103.4079} {arXiv:1103.4079 [hep-th]} \BibitemShut
  {NoStop}%
\bibitem [{\citenamefont {Jiang}\ \emph {et~al.}(2020)\citenamefont {Jiang},
  \citenamefont {Komatsu},\ and\ \citenamefont {Vescovi}}]{Jiang:2019xdz}%
  \BibitemOpen
  \bibfield  {author} {\bibinfo {author} {\bibfnamefont {Yunfeng}\ \bibnamefont
  {Jiang}}, \bibinfo {author} {\bibfnamefont {Shota}\ \bibnamefont {Komatsu}},
  \ and\ \bibinfo {author} {\bibfnamefont {Edoardo}\ \bibnamefont {Vescovi}},\
  }\bibfield  {title} {\enquote {\bibinfo {title} {{Structure constants in $
  \mathcal{N} $ = 4 SYM at finite coupling as worldsheet g-function}},}\ }\href
  {\doibase 10.1007/JHEP07(2020)037} {\bibfield  {journal} {\bibinfo  {journal}
  {JHEP}\ }\textbf {\bibinfo {volume} {07}},\ \bibinfo {pages} {037} (\bibinfo
  {year} {2020})},\ \Eprint {http://arxiv.org/abs/1906.07733} {arXiv:1906.07733
  [hep-th]} \BibitemShut {NoStop}%
\bibitem [{\citenamefont {Jiang}\ \emph {et~al.}(2019)\citenamefont {Jiang},
  \citenamefont {Komatsu},\ and\ \citenamefont {Vescovi}}]{Jiang:2019zig}%
  \BibitemOpen
  \bibfield  {author} {\bibinfo {author} {\bibfnamefont {Yunfeng}\ \bibnamefont
  {Jiang}}, \bibinfo {author} {\bibfnamefont {Shota}\ \bibnamefont {Komatsu}},
  \ and\ \bibinfo {author} {\bibfnamefont {Edoardo}\ \bibnamefont {Vescovi}},\
  }\bibfield  {title} {\enquote {\bibinfo {title} {{Exact Three-Point Functions
  of Determinant Operators in Planar $N=4$ Supersymmetric Yang-Mills
  Theory}},}\ }\href {\doibase 10.1103/PhysRevLett.123.191601} {\bibfield
  {journal} {\bibinfo  {journal} {Phys. Rev. Lett.}\ }\textbf {\bibinfo
  {volume} {123}},\ \bibinfo {pages} {191601} (\bibinfo {year} {2019})},\
  \Eprint {http://arxiv.org/abs/1907.11242} {arXiv:1907.11242 [hep-th]}
  \BibitemShut {NoStop}%
\bibitem [{\citenamefont {Holguin}\ and\ \citenamefont
  {Weng}(2023)}]{Holguin:2022zii}%
  \BibitemOpen
  \bibfield  {author} {\bibinfo {author} {\bibfnamefont {Adolfo}\ \bibnamefont
  {Holguin}}\ and\ \bibinfo {author} {\bibfnamefont {Wayne~W.}\ \bibnamefont
  {Weng}},\ }\bibfield  {title} {\enquote {\bibinfo {title} {{Orbit averaging
  coherent states: holographic three-point functions of AdS giant
  gravitons}},}\ }\href {\doibase 10.1007/JHEP05(2023)167} {\bibfield
  {journal} {\bibinfo  {journal} {JHEP}\ }\textbf {\bibinfo {volume} {05}},\
  \bibinfo {pages} {167} (\bibinfo {year} {2023})},\ \Eprint
  {http://arxiv.org/abs/2211.03805} {arXiv:2211.03805 [hep-th]} \BibitemShut
  {NoStop}%
\bibitem [{\citenamefont {Wu}\ \emph {et~al.}(2026)\citenamefont {Wu},
  \citenamefont {Jiang}, \citenamefont {Liu},\ and\ \citenamefont
  {Zhang}}]{Wu:2025ott}%
  \BibitemOpen
  \bibfield  {author} {\bibinfo {author} {\bibfnamefont {Yu}~\bibnamefont
  {Wu}}, \bibinfo {author} {\bibfnamefont {Yunfeng}\ \bibnamefont {Jiang}},
  \bibinfo {author} {\bibfnamefont {Chang}\ \bibnamefont {Liu}}, \ and\
  \bibinfo {author} {\bibfnamefont {Yang}\ \bibnamefont {Zhang}},\ }\bibfield
  {title} {\enquote {\bibinfo {title} {{All giant graviton two-point functions
  at two-loops}},}\ }\href {\doibase 10.1007/JHEP03(2026)097} {\bibfield
  {journal} {\bibinfo  {journal} {JHEP}\ }\textbf {\bibinfo {volume} {03}},\
  \bibinfo {pages} {097} (\bibinfo {year} {2026})},\ \Eprint
  {http://arxiv.org/abs/2509.23161} {arXiv:2509.23161 [hep-th]} \BibitemShut
  {NoStop}%
\bibitem [{\citenamefont {Jiang}\ \emph {et~al.}(2024)\citenamefont {Jiang},
  \citenamefont {Wu},\ and\ \citenamefont {Zhang}}]{Jiang:2023uut}%
  \BibitemOpen
  \bibfield  {author} {\bibinfo {author} {\bibfnamefont {Yunfeng}\ \bibnamefont
  {Jiang}}, \bibinfo {author} {\bibfnamefont {Yu}~\bibnamefont {Wu}}, \ and\
  \bibinfo {author} {\bibfnamefont {Yang}\ \bibnamefont {Zhang}},\ }\bibfield
  {title} {\enquote {\bibinfo {title} {{Giant correlators at quantum level}},}\
  }\href {\doibase 10.1007/JHEP05(2024)345} {\bibfield  {journal} {\bibinfo
  {journal} {JHEP}\ }\textbf {\bibinfo {volume} {05}},\ \bibinfo {pages} {345}
  (\bibinfo {year} {2024})},\ \Eprint {http://arxiv.org/abs/2311.16791}
  {arXiv:2311.16791 [hep-th]} \BibitemShut {NoStop}%
\bibitem [{\citenamefont {Chen}\ \emph {et~al.}(2025)\citenamefont {Chen},
  \citenamefont {Jiang},\ and\ \citenamefont {Zhou}}]{Chen:2025yxg}%
  \BibitemOpen
  \bibfield  {author} {\bibinfo {author} {\bibfnamefont {Junding}\ \bibnamefont
  {Chen}}, \bibinfo {author} {\bibfnamefont {Yunfeng}\ \bibnamefont {Jiang}}, \
  and\ \bibinfo {author} {\bibfnamefont {Xinan}\ \bibnamefont {Zhou}},\
  }\bibfield  {title} {\enquote {\bibinfo {title} {{Giant Graviton Correlators
  as Defect Systems}},}\ }\href {\doibase 10.1103/hg9p-hblr} {\bibfield
  {journal} {\bibinfo  {journal} {Phys. Rev. Lett.}\ }\textbf {\bibinfo
  {volume} {135}},\ \bibinfo {pages} {081602} (\bibinfo {year} {2025})},\
  \Eprint {http://arxiv.org/abs/2503.22987} {arXiv:2503.22987 [hep-th]}
  \BibitemShut {NoStop}%
\bibitem [{\citenamefont {Chen}\ \emph {et~al.}(2026)\citenamefont {Chen},
  \citenamefont {Jiang},\ and\ \citenamefont {Zhou}}]{Chen:2026ium}%
  \BibitemOpen
  \bibfield  {author} {\bibinfo {author} {\bibfnamefont {Junding}\ \bibnamefont
  {Chen}}, \bibinfo {author} {\bibfnamefont {Yunfeng}\ \bibnamefont {Jiang}}, \
  and\ \bibinfo {author} {\bibfnamefont {Xinan}\ \bibnamefont {Zhou}},\
  }\bibfield  {title} {\enquote {\bibinfo {title} {{Defect approach to giant
  graviton dynamics}},}\ }\href {\doibase 10.1088/1361-6633/ae79a4} {\bibfield
  {journal} {\bibinfo  {journal} {Rept. Prog. Phys.}\ }\textbf {\bibinfo
  {volume} {89}},\ \bibinfo {pages} {067801} (\bibinfo {year} {2026})},\
  \Eprint {http://arxiv.org/abs/2602.13570} {arXiv:2602.13570 [hep-th]}
  \BibitemShut {NoStop}%
\bibitem [{\citenamefont {Brown}\ \emph
  {et~al.}(2024{\natexlab{a}})\citenamefont {Brown}, \citenamefont {Galvagno},\
  and\ \citenamefont {Wen}}]{Brown:2024tru}%
  \BibitemOpen
  \bibfield  {author} {\bibinfo {author} {\bibfnamefont {Augustus}\
  \bibnamefont {Brown}}, \bibinfo {author} {\bibfnamefont {Francesco}\
  \bibnamefont {Galvagno}}, \ and\ \bibinfo {author} {\bibfnamefont {Congkao}\
  \bibnamefont {Wen}},\ }\bibfield  {title} {\enquote {\bibinfo {title} {{Exact
  results for giant graviton four-point correlators}},}\ }\href {\doibase
  10.1007/JHEP07(2024)049} {\bibfield  {journal} {\bibinfo  {journal} {JHEP}\
  }\textbf {\bibinfo {volume} {07}},\ \bibinfo {pages} {049} (\bibinfo {year}
  {2024}{\natexlab{a}})},\ \Eprint {http://arxiv.org/abs/2403.17263}
  {arXiv:2403.17263 [hep-th]} \BibitemShut {NoStop}%
\bibitem [{\citenamefont {Dorigoni}\ \emph {et~al.}(2023)\citenamefont
  {Dorigoni}, \citenamefont {Green}, \citenamefont {Wen},\ and\ \citenamefont
  {Xie}}]{Dorigoni:2022cua}%
  \BibitemOpen
  \bibfield  {author} {\bibinfo {author} {\bibfnamefont {Daniele}\ \bibnamefont
  {Dorigoni}}, \bibinfo {author} {\bibfnamefont {Michael~B.}\ \bibnamefont
  {Green}}, \bibinfo {author} {\bibfnamefont {Congkao}\ \bibnamefont {Wen}}, \
  and\ \bibinfo {author} {\bibfnamefont {Haitian}\ \bibnamefont {Xie}},\
  }\bibfield  {title} {\enquote {\bibinfo {title} {{Modular-invariant large-N
  completion of an integrated correlator in $ \mathcal{N} $ = 4 supersymmetric
  Yang-Mills theory}},}\ }\href {\doibase 10.1007/JHEP04(2023)114} {\bibfield
  {journal} {\bibinfo  {journal} {JHEP}\ }\textbf {\bibinfo {volume} {04}},\
  \bibinfo {pages} {114} (\bibinfo {year} {2023})},\ \Eprint
  {http://arxiv.org/abs/2210.14038} {arXiv:2210.14038 [hep-th]} \BibitemShut
  {NoStop}%
\bibitem [{\citenamefont {Eden}\ \emph {et~al.}(2001)\citenamefont {Eden},
  \citenamefont {Petkou}, \citenamefont {Schubert},\ and\ \citenamefont
  {Sokatchev}}]{Eden:2000bk}%
  \BibitemOpen
  \bibfield  {author} {\bibinfo {author} {\bibfnamefont {Burkhard}\
  \bibnamefont {Eden}}, \bibinfo {author} {\bibfnamefont {Anastasios~C.}\
  \bibnamefont {Petkou}}, \bibinfo {author} {\bibfnamefont {Christian}\
  \bibnamefont {Schubert}}, \ and\ \bibinfo {author} {\bibfnamefont {Emery}\
  \bibnamefont {Sokatchev}},\ }\bibfield  {title} {\enquote {\bibinfo {title}
  {{Partial nonrenormalization of the stress tensor four point function in N=4
  SYM and AdS / CFT}},}\ }\href {\doibase 10.1016/S0550-3213(01)00151-1}
  {\bibfield  {journal} {\bibinfo  {journal} {Nucl. Phys. B}\ }\textbf
  {\bibinfo {volume} {607}},\ \bibinfo {pages} {191--212} (\bibinfo {year}
  {2001})},\ \Eprint {http://arxiv.org/abs/hep-th/0009106}
  {arXiv:hep-th/0009106} \BibitemShut {NoStop}%
\bibitem [{\citenamefont {Nirschl}\ and\ \citenamefont
  {Osborn}(2005)}]{Nirschl:2004pa}%
  \BibitemOpen
  \bibfield  {author} {\bibinfo {author} {\bibfnamefont {M.}~\bibnamefont
  {Nirschl}}\ and\ \bibinfo {author} {\bibfnamefont {H.}~\bibnamefont
  {Osborn}},\ }\bibfield  {title} {\enquote {\bibinfo {title} {{Superconformal
  Ward identities and their solution}},}\ }\href {\doibase
  10.1016/j.nuclphysb.2005.01.013} {\bibfield  {journal} {\bibinfo  {journal}
  {Nucl. Phys. B}\ }\textbf {\bibinfo {volume} {711}},\ \bibinfo {pages}
  {409--479} (\bibinfo {year} {2005})},\ \Eprint
  {http://arxiv.org/abs/hep-th/0407060} {arXiv:hep-th/0407060} \BibitemShut
  {NoStop}%
\bibitem [{\citenamefont {Pestun}(2012)}]{Pestun:2007rz}%
  \BibitemOpen
  \bibfield  {author} {\bibinfo {author} {\bibfnamefont {Vasily}\ \bibnamefont
  {Pestun}},\ }\bibfield  {title} {\enquote {\bibinfo {title} {{Localization of
  gauge theory on a four-sphere and supersymmetric Wilson loops}},}\ }\href
  {\doibase 10.1007/s00220-012-1485-0} {\bibfield  {journal} {\bibinfo
  {journal} {Commun. Math. Phys.}\ }\textbf {\bibinfo {volume} {313}},\
  \bibinfo {pages} {71--129} (\bibinfo {year} {2012})},\ \Eprint
  {http://arxiv.org/abs/0712.2824} {arXiv:0712.2824 [hep-th]} \BibitemShut
  {NoStop}%
\bibitem [{\citenamefont {Gerchkovitz}\ \emph {et~al.}(2017)\citenamefont
  {Gerchkovitz}, \citenamefont {Gomis}, \citenamefont {Ishtiaque},
  \citenamefont {Karasik}, \citenamefont {Komargodski},\ and\ \citenamefont
  {Pufu}}]{Gerchkovitz:2016gxx}%
  \BibitemOpen
  \bibfield  {author} {\bibinfo {author} {\bibfnamefont {Efrat}\ \bibnamefont
  {Gerchkovitz}}, \bibinfo {author} {\bibfnamefont {Jaume}\ \bibnamefont
  {Gomis}}, \bibinfo {author} {\bibfnamefont {Nafiz}\ \bibnamefont
  {Ishtiaque}}, \bibinfo {author} {\bibfnamefont {Avner}\ \bibnamefont
  {Karasik}}, \bibinfo {author} {\bibfnamefont {Zohar}\ \bibnamefont
  {Komargodski}}, \ and\ \bibinfo {author} {\bibfnamefont {Silviu~S.}\
  \bibnamefont {Pufu}},\ }\bibfield  {title} {\enquote {\bibinfo {title}
  {{Correlation Functions of Coulomb Branch Operators}},}\ }\href {\doibase
  10.1007/JHEP01(2017)103} {\bibfield  {journal} {\bibinfo  {journal} {JHEP}\
  }\textbf {\bibinfo {volume} {01}},\ \bibinfo {pages} {103} (\bibinfo {year}
  {2017})},\ \Eprint {http://arxiv.org/abs/1602.05971} {arXiv:1602.05971
  [hep-th]} \BibitemShut {NoStop}%
\bibitem [{\citenamefont {Nekrasov}(2003)}]{Nekrasov:2002qd}%
  \BibitemOpen
  \bibfield  {author} {\bibinfo {author} {\bibfnamefont {Nikita~A.}\
  \bibnamefont {Nekrasov}},\ }\bibfield  {title} {\enquote {\bibinfo {title}
  {{Seiberg-Witten prepotential from instanton counting}},}\ }\href {\doibase
  10.4310/ATMP.2003.v7.n5.a4} {\bibfield  {journal} {\bibinfo  {journal} {Adv.
  Theor. Math. Phys.}\ }\textbf {\bibinfo {volume} {7}},\ \bibinfo {pages}
  {831--864} (\bibinfo {year} {2003})},\ \Eprint
  {http://arxiv.org/abs/hep-th/0206161} {arXiv:hep-th/0206161} \BibitemShut
  {NoStop}%
\bibitem [{\citenamefont {Gottsche}\ \emph {et~al.}(2008)\citenamefont
  {Gottsche}, \citenamefont {Nakajima},\ and\ \citenamefont
  {Yoshioka}}]{Gottsche:2006tn}%
  \BibitemOpen
  \bibfield  {author} {\bibinfo {author} {\bibfnamefont {Lothar}\ \bibnamefont
  {Gottsche}}, \bibinfo {author} {\bibfnamefont {Hiraku}\ \bibnamefont
  {Nakajima}}, \ and\ \bibinfo {author} {\bibfnamefont {Kota}\ \bibnamefont
  {Yoshioka}},\ }\bibfield  {title} {\enquote {\bibinfo {title} {{Instanton
  counting and Donaldson invariants}},}\ }\href@noop {} {\bibfield  {journal}
  {\bibinfo  {journal} {J. Diff. Geom.}\ }\textbf {\bibinfo {volume} {80}},\
  \bibinfo {pages} {343--390} (\bibinfo {year} {2008})},\ \Eprint
  {http://arxiv.org/abs/math/0606180} {arXiv:math/0606180} \BibitemShut
  {NoStop}%
\bibitem [{\citenamefont {Bonelli}\ \emph {et~al.}(2014)\citenamefont
  {Bonelli}, \citenamefont {Sciarappa}, \citenamefont {Tanzini},\ and\
  \citenamefont {Vasko}}]{Bonelli:2014iza}%
  \BibitemOpen
  \bibfield  {author} {\bibinfo {author} {\bibfnamefont {Giulio}\ \bibnamefont
  {Bonelli}}, \bibinfo {author} {\bibfnamefont {Antonio}\ \bibnamefont
  {Sciarappa}}, \bibinfo {author} {\bibfnamefont {Alessandro}\ \bibnamefont
  {Tanzini}}, \ and\ \bibinfo {author} {\bibfnamefont {Petr}\ \bibnamefont
  {Vasko}},\ }\bibfield  {title} {\enquote {\bibinfo {title} {{Six-dimensional
  supersymmetric gauge theories, quantum cohomology of instanton moduli spaces
  and gl(N) Quantum Intermediate Long Wave Hydrodynamics}},}\ }\href {\doibase
  10.1007/JHEP07(2014)141} {\bibfield  {journal} {\bibinfo  {journal} {JHEP}\
  }\textbf {\bibinfo {volume} {07}},\ \bibinfo {pages} {141} (\bibinfo {year}
  {2014})},\ \Eprint {http://arxiv.org/abs/1403.6454} {arXiv:1403.6454
  [hep-th]} \BibitemShut {NoStop}%
\bibitem [{\citenamefont {Montonen}\ and\ \citenamefont
  {Olive}(1977)}]{Montonen:1977sn}%
  \BibitemOpen
  \bibfield  {author} {\bibinfo {author} {\bibfnamefont {C.}~\bibnamefont
  {Montonen}}\ and\ \bibinfo {author} {\bibfnamefont {David~I.}\ \bibnamefont
  {Olive}},\ }\bibfield  {title} {\enquote {\bibinfo {title} {{Magnetic
  Monopoles as Gauge Particles?}}}\ }\href {\doibase
  10.1016/0370-2693(77)90076-4} {\bibfield  {journal} {\bibinfo  {journal}
  {Phys. Lett. B}\ }\textbf {\bibinfo {volume} {72}},\ \bibinfo {pages}
  {117--120} (\bibinfo {year} {1977})}\BibitemShut {NoStop}%
\bibitem [{\citenamefont {Collier}\ and\ \citenamefont
  {Perlmutter}(2022)}]{Collier:2022emf}%
  \BibitemOpen
  \bibfield  {author} {\bibinfo {author} {\bibfnamefont {Scott}\ \bibnamefont
  {Collier}}\ and\ \bibinfo {author} {\bibfnamefont {Eric}\ \bibnamefont
  {Perlmutter}},\ }\bibfield  {title} {\enquote {\bibinfo {title} {{Harnessing
  S-duality in $ \mathcal{N} $ = 4 SYM {\&} supergravity as SL(2,
  {\ensuremath{\mathbb{Z}}})-averaged strings}},}\ }\href {\doibase
  10.1007/JHEP08(2022)195} {\bibfield  {journal} {\bibinfo  {journal} {JHEP}\
  }\textbf {\bibinfo {volume} {08}},\ \bibinfo {pages} {195} (\bibinfo {year}
  {2022})},\ \Eprint {http://arxiv.org/abs/2201.05093} {arXiv:2201.05093
  [hep-th]} \BibitemShut {NoStop}%
\bibitem [{\citenamefont {Paul}\ \emph {et~al.}(2023)\citenamefont {Paul},
  \citenamefont {Perlmutter},\ and\ \citenamefont {Raj}}]{Paul:2022piq}%
  \BibitemOpen
  \bibfield  {author} {\bibinfo {author} {\bibfnamefont {Hynek}\ \bibnamefont
  {Paul}}, \bibinfo {author} {\bibfnamefont {Eric}\ \bibnamefont {Perlmutter}},
  \ and\ \bibinfo {author} {\bibfnamefont {Himanshu}\ \bibnamefont {Raj}},\
  }\bibfield  {title} {\enquote {\bibinfo {title} {{Integrated correlators in $
  \mathcal{N} $ = 4 SYM via SL(2, {\ensuremath{\mathbb{Z}}}) spectral
  theory}},}\ }\href {\doibase 10.1007/JHEP01(2023)149} {\bibfield  {journal}
  {\bibinfo  {journal} {JHEP}\ }\textbf {\bibinfo {volume} {01}},\ \bibinfo
  {pages} {149} (\bibinfo {year} {2023})},\ \Eprint
  {http://arxiv.org/abs/2209.06639} {arXiv:2209.06639 [hep-th]} \BibitemShut
  {NoStop}%
\bibitem [{Note1()}]{Note1}%
  \BibitemOpen
  \bibinfo {note} {{It can be seen from the matrix model through the small
  coupling expansion of $Z_{1-\protect \text {loop}}$ that the $O(\lambda ^1)$
  term contains only ${\protect \rm tr}\protect \, a^2$, which implies that
  this contribution can be obtained by $\tau _2 \partial _{\tau _2}$ acting on
  the two-point function of giant gravitons (then normalized by the two-point
  function itself), so the complicated color factors arising from giant
  gravitons cancel out. Therefore there is no exponentially suppressed
  contribution for the $\zeta (3)$ term in \protect \eqref
  {eq:pert-2loop}.}}\BibitemShut {Stop}%
\bibitem [{\citenamefont {Billo}\ \emph {et~al.}(2018)\citenamefont {Billo},
  \citenamefont {Fucito}, \citenamefont {Lerda}, \citenamefont {Morales},
  \citenamefont {Stanev},\ and\ \citenamefont {Wen}}]{Billo:2017glv}%
  \BibitemOpen
  \bibfield  {author} {\bibinfo {author} {\bibfnamefont {M.}~\bibnamefont
  {Billo}}, \bibinfo {author} {\bibfnamefont {F.}~\bibnamefont {Fucito}},
  \bibinfo {author} {\bibfnamefont {A.}~\bibnamefont {Lerda}}, \bibinfo
  {author} {\bibfnamefont {J.~F.}\ \bibnamefont {Morales}}, \bibinfo {author}
  {\bibfnamefont {Ya.~S.}\ \bibnamefont {Stanev}}, \ and\ \bibinfo {author}
  {\bibfnamefont {Congkao}\ \bibnamefont {Wen}},\ }\bibfield  {title} {\enquote
  {\bibinfo {title} {{Two-point correlators in $N =2$ gauge theories}},}\
  }\href {\doibase 10.1016/j.nuclphysb.2017.11.003} {\bibfield  {journal}
  {\bibinfo  {journal} {Nucl. Phys. B}\ }\textbf {\bibinfo {volume} {926}},\
  \bibinfo {pages} {427--466} (\bibinfo {year} {2018})},\ \Eprint
  {http://arxiv.org/abs/1705.02909} {arXiv:1705.02909 [hep-th]} \BibitemShut
  {NoStop}%
\bibitem [{Note2()}]{Note2}%
  \BibitemOpen
  \bibinfo {note} {See \cite {Caron-Huot:2023wdh} for the study of determinant
  operators related to the one here considered at finite $N$ from a twistor
  space approach.}\BibitemShut {Stop}%
\bibitem [{\citenamefont {Eden}\ \emph
  {et~al.}(2012{\natexlab{a}})\citenamefont {Eden}, \citenamefont {Heslop},
  \citenamefont {Korchemsky},\ and\ \citenamefont {Sokatchev}}]{Eden:2011we}%
  \BibitemOpen
  \bibfield  {author} {\bibinfo {author} {\bibfnamefont {Burkhard}\
  \bibnamefont {Eden}}, \bibinfo {author} {\bibfnamefont {Paul}\ \bibnamefont
  {Heslop}}, \bibinfo {author} {\bibfnamefont {Gregory~P.}\ \bibnamefont
  {Korchemsky}}, \ and\ \bibinfo {author} {\bibfnamefont {Emery}\ \bibnamefont
  {Sokatchev}},\ }\bibfield  {title} {\enquote {\bibinfo {title} {{Hidden
  symmetry of four-point correlation functions and amplitudes in N=4 SYM}},}\
  }\href {\doibase 10.1016/j.nuclphysb.2012.04.007} {\bibfield  {journal}
  {\bibinfo  {journal} {Nucl. Phys. B}\ }\textbf {\bibinfo {volume} {862}},\
  \bibinfo {pages} {193--231} (\bibinfo {year} {2012}{\natexlab{a}})},\ \Eprint
  {http://arxiv.org/abs/1108.3557} {arXiv:1108.3557 [hep-th]} \BibitemShut
  {NoStop}%
\bibitem [{\citenamefont {Eden}\ \emph
  {et~al.}(2012{\natexlab{b}})\citenamefont {Eden}, \citenamefont {Heslop},
  \citenamefont {Korchemsky},\ and\ \citenamefont {Sokatchev}}]{Eden:2012tu}%
  \BibitemOpen
  \bibfield  {author} {\bibinfo {author} {\bibfnamefont {Burkhard}\
  \bibnamefont {Eden}}, \bibinfo {author} {\bibfnamefont {Paul}\ \bibnamefont
  {Heslop}}, \bibinfo {author} {\bibfnamefont {Gregory~P.}\ \bibnamefont
  {Korchemsky}}, \ and\ \bibinfo {author} {\bibfnamefont {Emery}\ \bibnamefont
  {Sokatchev}},\ }\bibfield  {title} {\enquote {\bibinfo {title} {{Constructing
  the correlation function of four stress-tensor multiplets and the
  four-particle amplitude in N=4 SYM}},}\ }\href {\doibase
  10.1016/j.nuclphysb.2012.04.013} {\bibfield  {journal} {\bibinfo  {journal}
  {Nucl. Phys. B}\ }\textbf {\bibinfo {volume} {862}},\ \bibinfo {pages}
  {450--503} (\bibinfo {year} {2012}{\natexlab{b}})},\ \Eprint
  {http://arxiv.org/abs/1201.5329} {arXiv:1201.5329 [hep-th]} \BibitemShut
  {NoStop}%
\bibitem [{\citenamefont {Chicherin}\ \emph {et~al.}(2016)\citenamefont
  {Chicherin}, \citenamefont {Drummond}, \citenamefont {Heslop},\ and\
  \citenamefont {Sokatchev}}]{Chicherin:2015edu}%
  \BibitemOpen
  \bibfield  {author} {\bibinfo {author} {\bibfnamefont {Dmitry}\ \bibnamefont
  {Chicherin}}, \bibinfo {author} {\bibfnamefont {James}\ \bibnamefont
  {Drummond}}, \bibinfo {author} {\bibfnamefont {Paul}\ \bibnamefont {Heslop}},
  \ and\ \bibinfo {author} {\bibfnamefont {Emery}\ \bibnamefont {Sokatchev}},\
  }\bibfield  {title} {\enquote {\bibinfo {title} {{All three-loop four-point
  correlators of half-BPS operators in planar $ \mathcal{N} $ = 4 SYM}},}\
  }\href {\doibase 10.1007/JHEP08(2016)053} {\bibfield  {journal} {\bibinfo
  {journal} {JHEP}\ }\textbf {\bibinfo {volume} {08}},\ \bibinfo {pages} {053}
  (\bibinfo {year} {2016})},\ \Eprint {http://arxiv.org/abs/1512.02926}
  {arXiv:1512.02926 [hep-th]} \BibitemShut {NoStop}%
\bibitem [{\citenamefont {Bachas}\ \emph {et~al.}(1999)\citenamefont {Bachas},
  \citenamefont {Bain},\ and\ \citenamefont {Green}}]{Bachas:1999um}%
  \BibitemOpen
  \bibfield  {author} {\bibinfo {author} {\bibfnamefont {Constantin~P.}\
  \bibnamefont {Bachas}}, \bibinfo {author} {\bibfnamefont {Pascal}\
  \bibnamefont {Bain}}, \ and\ \bibinfo {author} {\bibfnamefont {Michael~B.}\
  \bibnamefont {Green}},\ }\bibfield  {title} {\enquote {\bibinfo {title}
  {{Curvature terms in D-brane actions and their M theory origin}},}\ }\href
  {\doibase 10.1088/1126-6708/1999/05/011} {\bibfield  {journal} {\bibinfo
  {journal} {JHEP}\ }\textbf {\bibinfo {volume} {05}},\ \bibinfo {pages} {011}
  (\bibinfo {year} {1999})},\ \Eprint {http://arxiv.org/abs/hep-th/9903210}
  {arXiv:hep-th/9903210} \BibitemShut {NoStop}%
\bibitem [{\citenamefont {Basu}(2008)}]{Basu:2008gt}%
  \BibitemOpen
  \bibfield  {author} {\bibinfo {author} {\bibfnamefont {Anirban}\ \bibnamefont
  {Basu}},\ }\bibfield  {title} {\enquote {\bibinfo {title} {{Constraining the
  D3-brane effective action}},}\ }\href {\doibase
  10.1088/1126-6708/2008/09/124} {\bibfield  {journal} {\bibinfo  {journal}
  {JHEP}\ }\textbf {\bibinfo {volume} {09}},\ \bibinfo {pages} {124} (\bibinfo
  {year} {2008})},\ \Eprint {http://arxiv.org/abs/0808.2060} {arXiv:0808.2060
  [hep-th]} \BibitemShut {NoStop}%
\bibitem [{\citenamefont {Lin}\ \emph {et~al.}(2015)\citenamefont {Lin},
  \citenamefont {Shao}, \citenamefont {Wang},\ and\ \citenamefont
  {Yin}}]{Lin:2015ixa}%
  \BibitemOpen
  \bibfield  {author} {\bibinfo {author} {\bibfnamefont {Ying-Hsuan}\
  \bibnamefont {Lin}}, \bibinfo {author} {\bibfnamefont {Shu-Heng}\
  \bibnamefont {Shao}}, \bibinfo {author} {\bibfnamefont {Yifan}\ \bibnamefont
  {Wang}}, \ and\ \bibinfo {author} {\bibfnamefont {Xi}~\bibnamefont {Yin}},\
  }\bibfield  {title} {\enquote {\bibinfo {title} {{Higher derivative couplings
  in theories with sixteen supersymmetries}},}\ }\href {\doibase
  10.1103/PhysRevD.92.125017} {\bibfield  {journal} {\bibinfo  {journal} {Phys.
  Rev. D}\ }\textbf {\bibinfo {volume} {92}},\ \bibinfo {pages} {125017}
  (\bibinfo {year} {2015})},\ \Eprint {http://arxiv.org/abs/1503.02077}
  {arXiv:1503.02077 [hep-th]} \BibitemShut {NoStop}%
\bibitem [{\citenamefont {Dorigoni}\ and\ \citenamefont
  {Treilis}(2024)}]{Dorigoni:2024dhy}%
  \BibitemOpen
  \bibfield  {author} {\bibinfo {author} {\bibfnamefont {Daniele}\ \bibnamefont
  {Dorigoni}}\ and\ \bibinfo {author} {\bibfnamefont {Rudolfs}\ \bibnamefont
  {Treilis}},\ }\bibfield  {title} {\enquote {\bibinfo {title} {{Large-N
  integrated correlators in $ \mathcal{N} $ = 4 SYM: when resurgence meets
  modularity}},}\ }\href {\doibase 10.1007/JHEP07(2024)235} {\bibfield
  {journal} {\bibinfo  {journal} {JHEP}\ }\textbf {\bibinfo {volume} {07}},\
  \bibinfo {pages} {235} (\bibinfo {year} {2024})},\ \Eprint
  {http://arxiv.org/abs/2405.10204} {arXiv:2405.10204 [hep-th]} \BibitemShut
  {NoStop}%
\bibitem [{\citenamefont {Luo}\ and\ \citenamefont {Wang}(2023)}]{Luo:2022tqy}%
  \BibitemOpen
  \bibfield  {author} {\bibinfo {author} {\bibfnamefont {Conghuan}\
  \bibnamefont {Luo}}\ and\ \bibinfo {author} {\bibfnamefont {Yifan}\
  \bibnamefont {Wang}},\ }\bibfield  {title} {\enquote {\bibinfo {title}
  {{Casimir energy and modularity in higher-dimensional conformal field
  theories}},}\ }\href {\doibase 10.1007/JHEP07(2023)028} {\bibfield  {journal}
  {\bibinfo  {journal} {JHEP}\ }\textbf {\bibinfo {volume} {07}},\ \bibinfo
  {pages} {028} (\bibinfo {year} {2023})},\ \Eprint
  {http://arxiv.org/abs/2212.14866} {arXiv:2212.14866 [hep-th]} \BibitemShut
  {NoStop}%
\bibitem [{\citenamefont {De~Lillo}\ \emph {et~al.}(2026)\citenamefont
  {De~Lillo}, \citenamefont {Duan}, \citenamefont {Frau}, \citenamefont
  {Galvagno}, \citenamefont {Lerda}, \citenamefont {Vallarino},\ and\
  \citenamefont {Wen}}]{DeLillo:2025stg}%
  \BibitemOpen
  \bibfield  {author} {\bibinfo {author} {\bibfnamefont {L.}~\bibnamefont
  {De~Lillo}}, \bibinfo {author} {\bibfnamefont {Z.}~\bibnamefont {Duan}},
  \bibinfo {author} {\bibfnamefont {M.}~\bibnamefont {Frau}}, \bibinfo {author}
  {\bibfnamefont {F.}~\bibnamefont {Galvagno}}, \bibinfo {author}
  {\bibfnamefont {A.}~\bibnamefont {Lerda}}, \bibinfo {author} {\bibfnamefont
  {P.}~\bibnamefont {Vallarino}}, \ and\ \bibinfo {author} {\bibfnamefont
  {C.}~\bibnamefont {Wen}},\ }\bibfield  {title} {\enquote {\bibinfo {title}
  {{$ \mathcal{N}=2 $ universality at strong coupling}},}\ }\href {\doibase
  10.1007/JHEP02(2026)019} {\bibfield  {journal} {\bibinfo  {journal} {JHEP}\
  }\textbf {\bibinfo {volume} {02}},\ \bibinfo {pages} {019} (\bibinfo {year}
  {2026})},\ \Eprint {http://arxiv.org/abs/2510.27594} {arXiv:2510.27594
  [hep-th]} \BibitemShut {NoStop}%
\bibitem [{Note3()}]{Note3}%
  \BibitemOpen
  \bibinfo {note} {{Preliminary calculations for general sphere giant gravitons
  with dimension $\alpha N$ and AdS giant gravitons with dimension $\beta N$
  suggest that the spectral overlap can again be separated into two parts for
  SU$(N)$ theory. The first part, valid at all orders in $1/N$, relates the
  sphere giant gravitons and AdS giant gravitons by $N\leftrightarrow -N$ at a
  given $s$, while the exponentially suppressed parts are not related to each
  other in this simple manner. For U$(N)$ theory, there are no $O(N^{-N})$
  constributions, and the spectral overlap for sphere giant gravitons is fully
  related to the AdS giant graviton spectral overlap by $N \leftrightarrow
  -N$.}}\BibitemShut {Stop}%
\bibitem [{\citenamefont {Chester}\ \emph {et~al.}(2025)\citenamefont
  {Chester}, \citenamefont {Ferrero},\ and\ \citenamefont
  {Pavarini}}]{Chester:2025ssu}%
  \BibitemOpen
  \bibfield  {author} {\bibinfo {author} {\bibfnamefont {Shai~M.}\ \bibnamefont
  {Chester}}, \bibinfo {author} {\bibfnamefont {Pietro}\ \bibnamefont
  {Ferrero}}, \ and\ \bibinfo {author} {\bibfnamefont {Daniele~R.}\
  \bibnamefont {Pavarini}},\ }\bibfield  {title} {\enquote {\bibinfo {title}
  {{Modular invariant gluon-graviton scattering in AdS at one loop}},}\ }\href
  {\doibase 10.1007/JHEP08(2025)208} {\bibfield  {journal} {\bibinfo  {journal}
  {JHEP}\ }\textbf {\bibinfo {volume} {08}},\ \bibinfo {pages} {208} (\bibinfo
  {year} {2025})},\ \Eprint {http://arxiv.org/abs/2504.10319} {arXiv:2504.10319
  [hep-th]} \BibitemShut {NoStop}%
\bibitem [{\citenamefont {Goddard}\ \emph {et~al.}(1977)\citenamefont
  {Goddard}, \citenamefont {Nuyts},\ and\ \citenamefont
  {Olive}}]{Goddard:1976qe}%
  \BibitemOpen
  \bibfield  {author} {\bibinfo {author} {\bibfnamefont {P.}~\bibnamefont
  {Goddard}}, \bibinfo {author} {\bibfnamefont {J.}~\bibnamefont {Nuyts}}, \
  and\ \bibinfo {author} {\bibfnamefont {David~I.}\ \bibnamefont {Olive}},\
  }\bibfield  {title} {\enquote {\bibinfo {title} {{Gauge Theories and Magnetic
  Charge}},}\ }\href {\doibase 10.1016/0550-3213(77)90221-8} {\bibfield
  {journal} {\bibinfo  {journal} {Nucl. Phys. B}\ }\textbf {\bibinfo {volume}
  {125}},\ \bibinfo {pages} {1--28} (\bibinfo {year} {1977})}\BibitemShut
  {NoStop}%
\bibitem [{\citenamefont {Dorigoni}\ \emph
  {et~al.}(2022{\natexlab{b}})\citenamefont {Dorigoni}, \citenamefont {Green},\
  and\ \citenamefont {Wen}}]{Dorigoni:2022zcr}%
  \BibitemOpen
  \bibfield  {author} {\bibinfo {author} {\bibfnamefont {Daniele}\ \bibnamefont
  {Dorigoni}}, \bibinfo {author} {\bibfnamefont {Michael~B.}\ \bibnamefont
  {Green}}, \ and\ \bibinfo {author} {\bibfnamefont {Congkao}\ \bibnamefont
  {Wen}},\ }\bibfield  {title} {\enquote {\bibinfo {title} {{Exact results for
  duality-covariant integrated correlators in $\mathcal{N}=4$ SYM with general
  classical gauge groups}},}\ }\href {\doibase 10.21468/SciPostPhys.13.4.092}
  {\bibfield  {journal} {\bibinfo  {journal} {SciPost Phys.}\ }\textbf
  {\bibinfo {volume} {13}},\ \bibinfo {pages} {092} (\bibinfo {year}
  {2022}{\natexlab{b}})},\ \Eprint {http://arxiv.org/abs/2202.05784}
  {arXiv:2202.05784 [hep-th]} \BibitemShut {NoStop}%
\bibitem [{\citenamefont {Dorigoni}\ and\ \citenamefont
  {Vallarino}(2023)}]{Dorigoni:2023ezg}%
  \BibitemOpen
  \bibfield  {author} {\bibinfo {author} {\bibfnamefont {Daniele}\ \bibnamefont
  {Dorigoni}}\ and\ \bibinfo {author} {\bibfnamefont {Paolo}\ \bibnamefont
  {Vallarino}},\ }\bibfield  {title} {\enquote {\bibinfo {title}
  {{Exceptionally simple integrated correlators in $ \mathcal{N} $ = 4
  supersymmetric Yang-Mills theory}},}\ }\href {\doibase
  10.1007/JHEP09(2023)203} {\bibfield  {journal} {\bibinfo  {journal} {JHEP}\
  }\textbf {\bibinfo {volume} {09}},\ \bibinfo {pages} {203} (\bibinfo {year}
  {2023})},\ \Eprint {http://arxiv.org/abs/2308.15252} {arXiv:2308.15252
  [hep-th]} \BibitemShut {NoStop}%
\bibitem [{\citenamefont {Dorigoni}(2019)}]{Dorigoni:2014hea}%
  \BibitemOpen
  \bibfield  {author} {\bibinfo {author} {\bibfnamefont {Daniele}\ \bibnamefont
  {Dorigoni}},\ }\bibfield  {title} {\enquote {\bibinfo {title} {{An
  Introduction to Resurgence, Trans-Series and Alien Calculus}},}\ }\href
  {\doibase 10.1016/j.aop.2019.167914} {\bibfield  {journal} {\bibinfo
  {journal} {Annals Phys.}\ }\textbf {\bibinfo {volume} {409}},\ \bibinfo
  {pages} {167914} (\bibinfo {year} {2019})},\ \Eprint
  {http://arxiv.org/abs/1411.3585} {arXiv:1411.3585 [hep-th]} \BibitemShut
  {NoStop}%
\bibitem [{\citenamefont {Usyukina}\ and\ \citenamefont
  {Davydychev}(1993)}]{Usyukina:1993ch}%
  \BibitemOpen
  \bibfield  {author} {\bibinfo {author} {\bibfnamefont {N.~I.}\ \bibnamefont
  {Usyukina}}\ and\ \bibinfo {author} {\bibfnamefont {Andrei~I.}\ \bibnamefont
  {Davydychev}},\ }\bibfield  {title} {\enquote {\bibinfo {title} {{Exact
  results for three and four point ladder diagrams with an arbitrary number of
  rungs}},}\ }\href {\doibase 10.1016/0370-2693(93)91118-7} {\bibfield
  {journal} {\bibinfo  {journal} {Phys. Lett. B}\ }\textbf {\bibinfo {volume}
  {305}},\ \bibinfo {pages} {136--143} (\bibinfo {year} {1993})}\BibitemShut
  {NoStop}%
\bibitem [{\citenamefont {Wen}\ and\ \citenamefont
  {Zhang}(2022)}]{Wen:2022oky}%
  \BibitemOpen
  \bibfield  {author} {\bibinfo {author} {\bibfnamefont {Congkao}\ \bibnamefont
  {Wen}}\ and\ \bibinfo {author} {\bibfnamefont {Shun-Qing}\ \bibnamefont
  {Zhang}},\ }\bibfield  {title} {\enquote {\bibinfo {title} {{Integrated
  correlators in $ \mathcal{N} $ = 4 super Yang-Mills and periods}},}\ }\href
  {\doibase 10.1007/JHEP05(2022)126} {\bibfield  {journal} {\bibinfo  {journal}
  {JHEP}\ }\textbf {\bibinfo {volume} {05}},\ \bibinfo {pages} {126} (\bibinfo
  {year} {2022})},\ \Eprint {http://arxiv.org/abs/2203.01890} {arXiv:2203.01890
  [hep-th]} \BibitemShut {NoStop}%
\bibitem [{\citenamefont {Brown}\ \emph
  {et~al.}(2024{\natexlab{b}})\citenamefont {Brown}, \citenamefont {Heslop},
  \citenamefont {Wen},\ and\ \citenamefont {Xie}}]{Brown:2023zbr}%
  \BibitemOpen
  \bibfield  {author} {\bibinfo {author} {\bibfnamefont {Augustus}\
  \bibnamefont {Brown}}, \bibinfo {author} {\bibfnamefont {Paul}\ \bibnamefont
  {Heslop}}, \bibinfo {author} {\bibfnamefont {Congkao}\ \bibnamefont {Wen}}, \
  and\ \bibinfo {author} {\bibfnamefont {Haitian}\ \bibnamefont {Xie}},\
  }\bibfield  {title} {\enquote {\bibinfo {title} {{Integrated correlators in
  $\mathcal{N}=4$ SYM beyond localisation}},}\ }\href {\doibase
  10.1103/PhysRevLett.132.101602} {\bibfield  {journal} {\bibinfo  {journal}
  {Phys. Rev. Lett.}\ }\textbf {\bibinfo {volume} {132}},\ \bibinfo {pages}
  {101602} (\bibinfo {year} {2024}{\natexlab{b}})},\ \Eprint
  {http://arxiv.org/abs/2308.07219} {arXiv:2308.07219 [hep-th]} \BibitemShut
  {NoStop}%
\bibitem [{\citenamefont {Belokurov}\ and\ \citenamefont
  {Usyukina}(1983)}]{Belokurov:1983km}%
  \BibitemOpen
  \bibfield  {author} {\bibinfo {author} {\bibfnamefont {V.~V.}\ \bibnamefont
  {Belokurov}}\ and\ \bibinfo {author} {\bibfnamefont {N.~I.}\ \bibnamefont
  {Usyukina}},\ }\bibfield  {title} {\enquote {\bibinfo {title} {{Calculation
  of ladder diagrams in arbitrary order}},}\ }\href {\doibase
  10.1088/0305-4470/16/12/026} {\bibfield  {journal} {\bibinfo  {journal} {J.
  Phys. A}\ }\textbf {\bibinfo {volume} {16}},\ \bibinfo {pages} {2811--2816}
  (\bibinfo {year} {1983})}\BibitemShut {NoStop}%
\bibitem [{\citenamefont {Fiamberti}\ \emph {et~al.}(2008)\citenamefont
  {Fiamberti}, \citenamefont {Santambrogio}, \citenamefont {Sieg},\ and\
  \citenamefont {Zanon}}]{Fiamberti:2007rj}%
  \BibitemOpen
  \bibfield  {author} {\bibinfo {author} {\bibfnamefont {F.}~\bibnamefont
  {Fiamberti}}, \bibinfo {author} {\bibfnamefont {A.}~\bibnamefont
  {Santambrogio}}, \bibinfo {author} {\bibfnamefont {C.}~\bibnamefont {Sieg}},
  \ and\ \bibinfo {author} {\bibfnamefont {D.}~\bibnamefont {Zanon}},\
  }\bibfield  {title} {\enquote {\bibinfo {title} {{Wrapping at four loops in
  N=4 SYM}},}\ }\href {\doibase 10.1016/j.physletb.2008.06.061} {\bibfield
  {journal} {\bibinfo  {journal} {Phys. Lett. B}\ }\textbf {\bibinfo {volume}
  {666}},\ \bibinfo {pages} {100--105} (\bibinfo {year} {2008})},\ \Eprint
  {http://arxiv.org/abs/0712.3522} {arXiv:0712.3522 [hep-th]} \BibitemShut
  {NoStop}%
\bibitem [{\citenamefont {Caron-Huot}\ \emph {et~al.}(2023)\citenamefont
  {Caron-Huot}, \citenamefont {Coronado},\ and\ \citenamefont
  {M{\"u}hlmann}}]{Caron-Huot:2023wdh}%
  \BibitemOpen
  \bibfield  {author} {\bibinfo {author} {\bibfnamefont {Simon}\ \bibnamefont
  {Caron-Huot}}, \bibinfo {author} {\bibfnamefont {Frank}\ \bibnamefont
  {Coronado}}, \ and\ \bibinfo {author} {\bibfnamefont {Beatrix}\ \bibnamefont
  {M{\"u}hlmann}},\ }\bibfield  {title} {\enquote {\bibinfo {title}
  {{Determinants in self-dual $ \mathcal{N} $ = 4 SYM and twistor space}},}\
  }\href {\doibase 10.1007/JHEP08(2023)008} {\bibfield  {journal} {\bibinfo
  {journal} {JHEP}\ }\textbf {\bibinfo {volume} {08}},\ \bibinfo {pages} {008}
  (\bibinfo {year} {2023})},\ \Eprint {http://arxiv.org/abs/2304.12341}
  {arXiv:2304.12341 [hep-th]} \BibitemShut {NoStop}%
\end{thebibliography}%

%\bibliography{ref-giant}

%\begin{thebibliography}{3}

\end{document}